\documentclass[onecolumn]{article}

\usepackage[margin=2.54cm]{geometry}
\usepackage{hyperref}
\usepackage{authblk}
\usepackage{abstract}
\usepackage{titlesec}
\usepackage{cuted}
\usepackage{enumitem}

\usepackage{amsmath,amsfonts,amssymb,amsthm,mathtools,physics}

\usepackage{subcaption}
\usepackage{tikz}
\usepackage{qcircuit}
\usepackage{framed}

\usepackage[boxed]{algorithm}
\usepackage{algorithmic}

\usepackage[backend=biber,citestyle=numeric-comp,bibstyle=numeric,sorting=none,sortcites=true,maxnames=5,minnames=4]{biblatex}
\DeclareFieldFormat{doi}{%
  \hfil\penalty90\hfilneg\space DOI\addcolon\addnbspace
  \ifhyperref
    {\href{https://doi.org/#1}{\nolinkurl{#1}}}
    {\nolinkurl{#1}}}

\newtheorem{Theorem}{Theorem}
\newtheorem{Corollary}{Corollary}
\newtheorem{Lemma}{Lemma} 
\newtheorem{Definition}{Definition}

\newcommand{\Herm}{\text{Herm}}
\newcommand{\Cl}{\mathcal{C}\ell}
\newcommand{\Sp}{\text{Sp}}
\newcommand{\Span}[1]{\ensuremath{ \langle #1 \rangle }}

\newcommand{\Z}{\mathbb{Z}}
\newcommand{\pr}{\ensuremath{\textup{pr}}}

\titleformat*{\section}{\normalsize\bfseries}
\titleformat*{\subsection}{\normalsize\bfseries}

\title{\large\textbf{Efficient classical simulation of quantum computation beyond Wigner positivity}}
\author[1,2]{Michael Zurel\footnote{Both authors contributed significantly.}}
\author[3]{Arne Heimendahl$^{*}$}
\affil[1]{Department of Physics and Astronomy, University of British Columbia, Vancouver, Canada}
\affil[2]{Stewart Blusson Quantum Matter Institute, University of British Columbia, Vancouver, Canada}
\affil[3]{Department of Mathematics and Computer Science, University of Cologne, Cologne, Germany}
\date{}

\begin{document}

\maketitle
\begin{abstract}
	\vspace{-8mm}
	We present the generalization of the CNC formalism, based on closed and noncontextual sets of Pauli observables, to the setting of odd-prime-dimensional qudits. By introducing new CNC-type phase space point operators, we construct a quasiprobability representation for quantum computation which is covariant with respect to the Clifford group and positivity preserving under Pauli measurements, and whose nonnegative sector strictly contains the subtheory of quantum theory described by nonnegative Wigner functions. This allows for a broader class of magic state quantum circuits to be efficiently classically simulated than those covered by the stabilizer formalism and Wigner function methods.
\end{abstract}

\section{Introduction}\label{Section:Introduction}

Quasiprobability representations have long played an important role in physics bridging the gap between classical and quantum. Originally conceived by Wigner~\cite{Wigner1932} as a means to adapt phase space methods of statistical mechanics to quantum physics, they have since found many applications in diverse contexts~\cite{Husimi1940,Stratonovich1956,Glauber1963,Sudarshan1963,BrifMann1999,KenfackZyczkowski2004,Ferrie2011}. Notably, they have proven useful in describing and understanding quantum computation~\cite{Galvao2005,CormickGalvaoGottesman2006,Gross2006,Gross20062,Gross2008,MariEisert2012,VeitchEmerson2012}.

Gross' Wigner function~\cite{Gross2006,Gross20062,Gross2008}---a Wigner function for systems of odd-dimensional qudits---has proven particularly effective for describing quantum computation. It defines a noncontextual (thus classical) model of the stabilizer subtheory~\cite{Spekkens2008,HowardEmerson2014,OkayRaussendorf2017,SchmidPusey2022,RaussendorfFeldmann2023}, a realm where the function remains nonnegative. Negativity in this function has been proposed as the source of quantum computational power~\cite{Galvao2005}. This aligns with the traditional view where, as negativity in the Wigner function is the feature that distinguishes it from a classical probability distribution over a phase space, it is considered an indicator of nonclassical behaviour~\cite{KenfackZyczkowski2004}.

Veitch et al~\cite{VeitchEmerson2012} showed that negativity in this Wigner function is a necessary condition for a quantum computational advantage in the model of quantum computation with magic states (QCM) on odd-prime dimensional qudits. This is achieved by introducing an efficient classical simulation algorithm for QCM circuits that applies whenever the Wigner representation of the input state of the quantum circuit in question is nonnegative. Further, negativity in the representation of the input state can be used to quantify the cost of classical simulation, since in that case another sampling-based classical simulation algorithm can be applied where the number of samples required to achieve a given error scales with the amount of negativity~\cite{PashayanBartlett2015}.

These result can be easily extended to quantum computation on qudits of any odd dimension~\cite{Zurel2020}, but issues arise when attempting to extend them to computation on even-dimensional qudits. It was widely believed that no Wigner function exists which describes quantum computation on even-dimensional qudits, and this was proven in a few special cases~\cite{Webb2016,Zhu2016,Zhu2017}. This issue has since been settled for the general case~\cite{SchmidPusey2022,RaussendorfFeldmann2023}, but in the meantime a range of alternative quasiprobability representations have been defined~\cite{DelfosseRaussendorf2015,HowardCampbell2017,KociaLove2017,RallKretschmer2019,DeBrotaStacey2020,RaussendorfZurel2020,SeddonCampbell2021}. These representations relax some of the assumptions that typically define a Wigner function~\cite{Stratonovich1956,BrifMann1999,RaussendorfFeldmann2023}, but they retain the properties required for classical simulation of quantum computation, namely, covariance with respect to the Clifford group and preservation of nonnegativity under Pauli measurements.

One of these recently introduced quasiprobability representations is the so-called CNC construction~\cite{RaussendorfZurel2020}. In the case of odd-dimensional qudits, phase space points of the Wigner function are identified with noncontextual value assignment functions on the set of Pauli observables~\cite{DelfosseRaussendorf2017}, however, in the case of even dimensions, such noncontextual value assignments do not exist~\cite{Mermin1993,OkayRaussendorf2017,RaussendorfFeldmann2023}. The CNC construction circumvents this issue by allowing the set of Pauli observables over which to value assignment is defined to also vary. That is, phase space points are identified with pairs $(\Omega,\gamma)$, where $\Omega$ is a noncontextual subset of the Pauli observables and $\gamma$ is a noncontextual value assignment on $\Omega$. ``CNC'' refers to these sets $\Omega$, CNC stands for Closed under inference and NonContextual, two technical constraints we impose on the sets that we will define later. The result is a much larger phase space than a typical Wigner function would have, as required for consistency with a proven memory lower bound for simulating quantum contextuality~\cite{KaranjaiBartlett2018}, but the properties of the phase space required for simulating quantum computations are retained. This construction applies to any Hilbert space dimension, though so far it has mostly been studied for systems of multiple qubits~\cite{RaussendorfZurel2020}.

In this paper, we address the CNC construction for the case of odd-prime dimensional qudits. In Ref.~\cite{RaussendorfZurel2020}, it was claimed that in this case the CNC construction is equivalent to Gross' Wigner function~\cite{Gross2006,Gross20062,Gross2008}, but this was later shown to be incorrect~\cite{RaussendorfZurel2020Erratum}. Here we show that it can actually outperform Gross' Wigner function. That is, although the CNC phase space contains all of the points of the Wigner function phase space, our approach also involves the introduction of new CNC-type phase space points that are not equivalent to Wigner function phase space points, or convex mixtures thereof. The resulting quasiprobability representation obtained by introducing these new phase space points remains closed under Clifford operations and Pauli measurements, ensuring that the representation is capable of simulating quantum computations in the magic state model.

We provide a characterization of the CNC-type phase space points for odd-prime-dimensional qudits. One of the primary differences between this case and the qubit case stems from the fact that there are no state-independent proofs of contextuality among Pauli observables on odd-dimensional qudits~\cite{OkayRaussendorf2017,RaussendorfFeldmann2023}. As a result, only the closure under inference condition is required for the characterization, not the noncontextuality condition. Despite this difference, the classification of CNC-type phase space point operators in odd-prime-dimensional qudits looks very similar to the qubit case.

We also compare this construction with the $\Lambda$ polytope model~\cite{ZurelRaussendorf2020,ZurelHeimendahl2024}, a recently introduced hidden variable model for QCM. In the multiqubit case, CNC-type phase space point operators are vertices of the $\Lambda$ polytopes, and so they are identified with some of the hidden variables of that model~\cite{Heimendahl2019,ZurelRaussendorf2020}. This is partially mirrored in the odd-prime-dimensional case where at least some of the CNC-type phase space point operators, including the phase space points of the Wigner function, are vertices of the qudit version of the $\Lambda$ polytope model~\cite{ZurelHeimendahl2024}.

The primary advantage of this new construction lies in the expanded set of states that can be positively represented by the new quasiprobability function. Specifically, our construction goes beyond the capabilities of the previous Wigner function~\cite{Gross2006,Gross20062,Gross2008,VeitchEmerson2012} and stabilizer~\cite{Gottesman1997,Gottesman1998,Gottesman1999,AaronsonGottesman2004} methods in identifying states that can be efficiently classically simulated. This expanded scope provides a more precise delineation of the classical-to-quantum transition in QCM on odd-prime-dimensional qudits, shedding light on the properties of magic states that enable a computational advantage in these systems and more generally offering new insights into the nature of computational resources in quantum computation with magic states.

The rest of this paper is organized as follows. In Section~\ref{Section:Background}, we provide necessary background on the model of quantum computation with magic states and quasiprobability representations of it. In Section~\ref{Section:PhaseSpaceDefinition}, we define the CNC phase space, and we determine its relation to the phase space of the Wigner function for odd-prime-dimensional qudits. We also consider a simplification that removes redundancy in the phase space points for the case of multiple qubits, and we show why this reduction fails for odd-prime-dimensional qudits, thus proving that the CNC construction does not reduce to the Wigner function in odd dimensions. In Section~\ref{Section:PhaseSpaceCharacterization}, we provide a characterization of the CNC phase space for odd-prime-dimensional qudits, this characterization partially mirrors the structure of the Wigner function phase space~\cite{DelfosseRaussendorf2017}, as well as that of the multiqubit CNC phase space~\cite[\S IV]{RaussendorfZurel2020}.

In Section~\ref{Section:ClassicalSimulation}, we demonstrate that this new phase space is closed under Clifford gates and Pauli measurements, and we show how these facts can be exploited to define a classical simulation algorithm for quantum computation. We also demonstrate that this algorithm is efficient whenever the function representing the input state is nonnegative and samples from this distribution can be obtained efficiently. In Section~\ref{Section:LambdaConnection}, we show how the CNC construction relates to the $\Lambda$-polytope model~\cite{ZurelRaussendorf2020,ZurelHeimendahl2024}, a probabilistic representation of quantum computation with magic states. Finally, in Section~\ref{Section:Conclusion}, we conclude with a discussion of the implications of our results for the classical-to-quantum transition in QCM and propose directions for future research.

\section{Background}\label{Section:Background}

In this section, we provide some background information. First, in Section~\ref{Section:MagicStates} we define the model of quantum computation that we are primarily working with, namely, quantum computation with magic states (QCM), and we introduce some necessary notation surrounding it. Then, in Section~\ref{Section:QuasiprobabilityRepresentations} we review previous quasiprobability representations for quantum computation, including the Wigner function for odd-prime-dimensional qudits.

\subsection{Quantum computation with magic states}\label{Section:MagicStates}

Quantum computation with magic states (QCM)~\cite{BravyiKitaev2005} is a universal model of quantum computation in which the allowed operations are restricted to the stabilizer operations---Clifford gates and Pauli measurements, possibly with classical side-processing and adaptivity. When acting on computational basis states, these operations are not universal for quantum computation, and any circuit consisting of only these operations can be efficiently simulated on a classical computer~\cite{Gottesman1998,Gottesman1999,AaronsonGottesman2004}. Normally, in the circuit model, universality requires additional operations such as $T$ gates~\cite{NielsenChuang2010}.

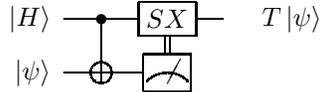
\begin{figure}
	\centering
	\mbox{
		\Qcircuit @C=1em @R=0.7em {
			\lstick{\ket{H}} & \ctrl{1} & \gate{SX} \cwx[1] & \qw & \rstick{T\ket{\psi}}\\
			\lstick{\ket{\psi}} & \targ & \meter
		}
	}
	\caption{Implementation of a $T$ gate by injection of a magic state~\cite[\S10.2.5]{NielsenChuang2010}. Only stabilizer operations (Clifford gates, Pauli measurements, and classical conditioning) are used, but by including the magic input state $\ket{H}:=\frac{1}{\sqrt{2}}\left(\ket{0}+e^{i\pi/4}\ket{1}\right)$, we can effectively implement a non-Clifford $T$ gate on the data qubit $\ket{\psi}$.\label{Figure:MagicTGate}}
\end{figure}

In the magic state model, universality is restored by the inclusion of additional ``magic'' input states. These are nonstabilizer states included as additional inputs to the quantum circuit, which allow for the implementation of nonstabilizer operations. See Figure~\ref{Figure:MagicTGate} for an example of the implementation of a non-Clifford $T$ gate by injection of a magic state, which proves universality of the model. Of course, without the capacity to perform non-Clifford gates, it's not obvious that these magic states can be prepared. Fortunately, magic state distillation protocols have been discovered, which consume many copies of noisy magic states and return fewer copies of the magic states with higher purity~\cite{BravyiKitaev2005}. This scheme is one of the leading candidates for scalable fault-tolerant quantum computation~\cite{CampbellVuillot2017}.

\subsubsection{Pauli-based quantum computation}

Another closely related model of computation is Pauli-based quantum computation (PBC)~\cite{BravyiSmolin2016,PeresGalvao2021,Peres2023}. This model is similar the magic state model, except that the operations are restricted even further without sacrificing computational universality. A Pauli-based quantum computation on $n$ qudits consists: (1)~preparation of an initial $n$-qudit magic state (an $n$-fold tensor product of a fixed single-qudit magic state suffices), followed by (2)~a sequence of pair-wise commuting Pauli measurements of length at most $n$.

Note that the measurement sequence is not necessarily determined in advance. Measurements can depend on the outcomes of previous measurements and classical randomness, and they are determined on the fly via classical side-processing. See Refs.~\cite{PeresGalvao2021,Peres2023} for more information on this model. Since this model can be obtained from the magic state model by imposing further restrictions on the operations and states that we allow, all of our results applying to the magic state model also apply to Pauli-based quantum computation.

\subsubsection{Notation for the stabilizer formalism}~\label{Section:StabilizerFormalism}

Here we introduce some notation surrounding the stabilizer formalism for qubits and for odd-prime-dimensional qudits. Most of the focus of this paper will be on systems of $n$ qudits, each with Hilbert space dimension $d$ where $d$ is an odd prime. But it will also be informative to compare against the case of multiple qubits ($d=2$), so in this section we introduce notation for qudits of any prime dimension $d$. The case of composite qudit Hilbert space dimension is somewhat more complicated. The difference between prime and composite dimensions is partially addressed elsewhere, for example see Refs.~\cite{Gross2006,ZurelHeimendahl2024,RaussendorfFeldmann2023,Gheorghiu2014}.

Up to overall phases, Pauli operators can be idenitided with elements of the vector space $\mathbb{Z}_d^{2n}$~\cite{Gross2006}. We fix a phase convention for the Pauli operators to be
\begin{equation}\label{eq:Paulis}
	\begin{cases}
		T_a=i^{-\langle a_z|a_x\rangle}\bigotimes\limits_{k=1}^nZ^{a_z[k]}X^{a_x[k]},\quad\text{if $d=2$,}\vspace{3mm}\\
		T_a=\omega^{-\langle a_z|a_x\rangle\cdot2^{-1}}\bigotimes\limits_{k=1}^nZ^{a_z[k]}X^{a_x[k]},\quad\text{if $d$ is an odd prime,}
	\end{cases}
\end{equation}
for each $a=(a_z,a_x)\in\mathbb{Z}_d^n\times\mathbb{Z}_d^n=:E$. Here $\omega:=\exp(2\pi i/d)$ is a primitive $d^{th}$ root of unity. The exponent in the first case, $-\braket{a_z}{a_x}$, is computed mod $4$, and in the second case, $-\braket{a_z}{a_x}\cdot2^{-1}$ is computed in $\mathbb{Z}_d$. The local Pauli operators are the $d$-dimensional generalization of the qubit Pauli operators, given by
\begin{equation}
	X=\sum\limits_{j\in\mathbb{Z}_d}\ket{j+1}\bra{j}\;\;\text{and}\;\; Z=\sum\limits_{j\in\mathbb{Z}_d}\omega^j\ket{j}\bra{j}.
\end{equation}
The symplectic product $[\cdot,\cdot]:E\times E\to\mathbb{Z}_d$, defined as $[a,b]:=\braket{a_z}{b_x}-\braket{a_x}{b_z}$, computes the commutator of the Pauli operators through
\begin{equation}
	[T_a,T_b]:=T_aT_bT_a^{-1}T_b^{-1}=\omega^{[a,b]},\quad\forall a,b\in E.
\end{equation}
When $[a,b]=0$, we say $a$ and $b$ are orthogonal with respect to the symplectic inner product. For orthogonal $a,b$, we sometimes also say $a$ and $b$ commute, since in this case the corresponding Pauli operators $T_a$ and $T_b$ commute. For a set of labels of Pauli operators $A\subset E$, the orthogonal complement of $A$, is the the set $A^\perp:=\{b\in E\;|\;[a,b]=0\;\forall a\in A\}$. I.e., this is the set of all labels of Pauli operators that commute with every Pauli operator in $A$. For an element $a\in E$, we will denote $a^\perp:=\{a\}^\perp$ the set of labels of Pauli operators that commute with $a$.

We define a $\mathbb{Z}_d$-valued function $\beta$ that tracks how commuting Pauli operators compose through the relation
\begin{equation}
	T_aT_b=\omega^{-\beta(a,b)}T_{a+b},\quad\forall a,b\in E\text{ such that }[a,b]=0.
\end{equation}
In the case where $d$ is odd, with the phase convention chosen in Eq.~\eqref{eq:Paulis}, $\beta(a,b)=0$ for all orthogonal $a,b\in E$. For $d=2$, no phase convention with this property exists~\cite{OkayRaussendorf2017,RaussendorfFeldmann2023}.

The Clifford group, $\Cl$, is a subgroup of the unitary group consisting of operators that map Pauli operators to Pauli operators (up to overall phases) under conjugation. The Clifford group acts on the Pauli operators by conjugation as
\begin{equation}\label{Equation:CliffordActionPaulis}
	gT_ag^\dagger=\omega^{\Phi_g(a)}T_{S_ga},\quad\forall g\in\Cl, a\in E,
\end{equation}
where $S_g\in\Sp(E)$ is a symplectic map on $E$~\cite{Gross2006}, and the function $\Phi_g:E\to\mathbb{Z}_d$, defined by this equation, tracks the extra phases on the Pauli operators that get picked up through this action.

A Pauli measurement is specified by a label $a\in E$ of a Pauli observable. The probability of obtaining measurement outcome $\omega^{r(a)}$ is given by the Born rule $P(r(a)|a)=\Tr(\rho\Pi_a^r)$ where
\begin{equation}
	\Pi_a^r:=\frac{1}{d}\sum\limits_{j\in\mathbb{Z}_d}\omega^{-jr(a)}T_a^j
\end{equation}
is the projector onto the eigenspace of $T_a$ corresponding to eigenvalue $\omega^{r(a)}, r(a)\in\mathbb{Z}_d$. More generally, we could measure any set of pair-wise commuting elements of $E$ simultaneously. In the symplectic vector space picture, this corresponds to measuring an isotropic subspace of $E$, i.e., a subspace $I$ of the vector space $E$ on which the symplectic form vanishes, $I\subset I^\perp$. Then the corresponding projector is
\begin{equation}
	\Pi_I^r:=\frac{1}{|I|}\sum\limits_{a\in I}\omega^{-r(a)}T_a
\end{equation}
where again the probability of obtaining the measurement outcomes $\left\{\omega^{r(a)}\;|\;a\in I\right\}$ is given by the Born rule, $P(r|I)=\Tr(\rho\Pi_I^r)$. For consistency, the measurement outcomes must satisfy
\begin{equation}
	\omega^{-r(a)-r(b)}T_aT_b=\omega^{-r(a+b)}T_{a+b},
\end{equation}
or, equivalently,
\begin{equation}
	r(a)+r(b)-r(a+b)=-\beta(a,b).
\end{equation}

\subsection{Quasiprobability representations}~\label{Section:QuasiprobabilityRepresentations}

The term quasiprobability representation means different things to different people. For example, some define quasiprobability representations as representations satisfying the Stratonowich-Weyl criteria~\cite{Stratonovich1956,BrifMann1999}, or finite-dimensional generalizations thereof~\cite{RaussendorfFeldmann2023}. Others admit only frame representations~\cite{FerrieEmerson2008,FerrieEmerson2009}. See, for example, Refs.~\cite{Ferrie2011,DavisGhose2021,RaussendorfFeldmann2023,SchmidSpekkens2024} for other types of quasiprobability representations in physics, and in particular, quasiprobability representations for finite-dimensional quantum theory.

Here we impose very few restrictions on the types of quasiprobability representations we allow. We impose only the constraints required for the representation to be useful for describing quantum computations. Denote by $\Herm(\mathcal{H})$ the real vector space of Hermitian operators on Hilbert space $\mathcal{H}$. We start by defining a finite set of operators $\{A_\alpha\;|\;\alpha\in\mathcal{V}\}$ on the $n$-qudit Hilbert space $(\mathbb{C}^d)^{\otimes n}\simeq\mathbb{C}^{d^n}$, with the following properties:
\begin{itemize}[leftmargin=20mm]
	\item[\hypertarget{QR1}{(QR1)}] $A_\alpha^\dagger=A_\alpha,\quad\forall\alpha\in\mathcal{V}$,
	\item[\hypertarget{QR2}{(QR2)}] $\Tr(A_\alpha)=1,\quad\forall\alpha\in\mathcal{V}$,
	\item[\hypertarget{QR3}{(QR3)}] $\text{Span}(\{A_\alpha\;|\;\alpha\in\mathcal{V}\})=\Herm(\mathcal{H})$.
\end{itemize}
As a result of property \hyperlink{QR3}{(QR3)}, any $n$-qudit quantum state represented by a density matrix $\rho$ can be expanded in these operators as
\begin{equation}\label{Equation:QRStates}
	\rho=\sum\limits_{\alpha\in\mathcal{V}}Q_\rho(\alpha)A_\alpha.
\end{equation}
From \hyperlink{QR1}{(QR1)}, the expansion coefficients are real for Hermitian operators, and with \hyperlink{QR2}{(QR2)}, taking a trace of this equation we obtain $1=\sum_{\alpha\in\mathcal{V}}Q_\rho(\alpha)$. This is the motivation for the term quasiprobability, the function $Q_\rho:\mathcal{V}\to\mathbb{R}$ looks like a probability distribution over the generalized phase space $\mathcal{V}$, except for the fact that it can take negative values. This function is how we represent states.

In order to describe quantum computations, we will be interested only in quasiprobability representation which are closed under the dynamics of quantum computation with magic states---Clifford gates and Pauli measurements. Therefore, we consider only quasiprobability representations satisfying the following additional constraints.
\begin{itemize}[leftmargin=20mm]
	\item[\hypertarget{QR4}{(QR4)}] For any Clifford group element $g\in\Cl$, $gA_\alpha g^\dagger=A_{g\cdot\alpha}$ with $g\cdot\alpha\in\mathcal{V}$
	\item[\hypertarget{QR5}{(QR5)}] For any Pauli measurement $a\in E$ and any measurement outcome $s\in\mathbb{Z}_d$, we have
	\begin{equation}
		\Pi_a^sA_\alpha\Pi_a^s=\sum\limits_{\beta\in\mathcal{V}}q_{\alpha,a}(\beta,s)A_\beta
	\end{equation}
	with $q_{\alpha,a}(\beta,s)\ge0,\;\forall\alpha,a,\beta,s$ and $\sum_{\beta,s}q_{\alpha,a}(\beta,s)=1$.
\end{itemize}
That is, according to \hyperlink{QR4}{(QR4)} Clifford operations are represented by a deterministic update map $\alpha\mapsto g\cdot\alpha$ on the phase space, and according to \hyperlink{QR5}{(QR5)} Pauli measurements are represented by a stochastic map $q_{\alpha,a}$. Once the operators $\{A_\alpha\}_{\alpha\in\mathcal{V}}$ are defined, the quasiprobability representation is determined, with $Q_\rho$ in Eq.~\eqref{Equation:QRStates} giving the representation of states, and \hyperlink{QR4}{(QR4)} and \hyperlink{QR5}{(QR5)} giving the representation of computational dynamics.

\subsubsection{The Wigner function}

Gross' discrete Wigner function~\cite{Gross2006,Gross20062,Gross2008} is a quasiprobability representation that has proven particularly effective at describing quantum computation on odd-dimensional qudits. The phase space of the Wigner function for odd-dimensional qudits is $V:=\mathbb{Z}_d^{2n}$, i.e., it is isomorphic to the labels of the Pauli operators. To each phase space point, we identify a phase space point operator defined as
\begin{equation}
	A_u=\frac{1}{d^n}\sum\limits_{b\in E}\omega^{[u,b]}T_b,\quad\forall u\in\mathbb{Z}_d^{2n}=:V.
\end{equation}
These operators are orthogonal, $\Tr(A_uA_v)=d^n\delta_{u,v}$, and they form a basis for $\Herm(\mathcal{H})$. The Wigner function of a state $\rho$ is defined as
\begin{equation}
	W_\rho(u)=\frac{1}{d^n}\Tr(\rho A_u),\quad\forall u\in V.
\end{equation}
These values are equal to the expansion coefficients in the expression $\rho=\sum_{u\in V}W_\rho(u)A_u$.

This representation satisfies \hyperlink{QR4}{(QR4)} and \hyperlink{QR5}{(QR5)}, in addition to \hyperlink{QR1}{(QR1)}--\hyperlink{QR3}{(QR3)}. In particular, the Clifford group acts on the phase space point operators by conjugation as
\begin{equation}
	gA_u g^\dagger=A_{S_gu+a_g}
\end{equation}
where $S_g\in\Sp(V)$ is a symplectic map on $V$, and $a_g\in \mathbb{Z}_d^{2n}$~\cite{Gross2006,Gross20062,Gross2008}. As a result, the Wigner function is Clifford covariant, $W_{g\rho g^\dagger}(S_gu+a_g)=W_\rho(u)$. For any Pauli measurement $a\in E$ and any measurement outcome $s\in\mathbb{Z}_d$, we have
\begin{equation}
	\begin{cases}
		\Tr(\Pi_a^sA_u)=1\text{ and }\Pi_a^sA_u\Pi_a^s=\frac{1}{|\Gamma_{a,u}|}\sum\limits_{w\in \Gamma_{a,u}}A_w,\quad\text{if $s=[a,u]$},\\
		\Tr(\Pi_a^sA_u)=0\text{ and }\Pi_a^sA_u\Pi_a^s=0,\quad\text{otherwise,}
	\end{cases}
\end{equation}
where $\Gamma_{a,u}=\{w\in E\;|\;[w,v]=[u,v]\;\forall v\in a^\perp\}$~\cite{Zurel2020}.

These updates under Clifford gates and Pauli measurements provide a classical simulation algorithm for quantum computation with magic states based on sampling, Algorithm~\ref{Algorithm:WignerAlgorithm}. Further, these state updates can be computed efficiently on a classical computer. Therefore, whenever the input state of a QCM circuit has a nonnegative representation, and this distribution can be efficiently sampled from, the entire computation can be efficiently simulated classically. This proves that negativity of the Wigner function is a necessary condition for a quantum computational advantage in QCM. This is a restatement of the main result from Ref.~\cite{VeitchEmerson2012}. For a more complete description of this formulation of the result, see Ref.~\cite[\S2]{Zurel2020}.

\begin{algorithm}
	\begin{algorithmic}[1]
		\REQUIRE $W_\rho(u)$; a Clifford+Pauli circuit $\mathcal{C}$
		\STATE sample a point $u\in V$ according to the probability distribution $W_\rho$
		\WHILE{end of circuit has not been reached}
			\IF{a Clifford gate $g\in\Cl$ is encountered}
				\STATE update $u\mapsto S_gu+a_g$
			\ENDIF
			\IF{a Pauli measurement $T_a,a\in E$ is encountered}
				\RETURN $\omega^{[a,u]}$ as the outcome of the measurement
				\STATE sample $w$ uniformly from $\Gamma_{a,u}$
				\STATE update $u\mapsto w$
			\ENDIF
		\ENDWHILE
	\end{algorithmic}
	\caption{A single run of the classical simulation algorithm for quantum computation with magic states based on sampling from the Wigner function. This algorithm applies only when the Wigner function of the input state is nonnegative. The algorithm provides samples from the joint probability distribution of the Pauli measurements in a quantum circuit consisting of Clifford gates and Pauli measurements applied to an input state $\rho$.\label{Algorithm:WignerAlgorithm}}
\end{algorithm}

\section{The CNC phase space}\label{Section:PhaseSpaceDefinition}

In this section we define a generalized phase space picture that can be used for describing quantum computation with magic states on qudits of any dimension. This phase space was first introduced in Ref.~\cite{RaussendorfZurel2020}, and there it was characterized for the case of qubits. In this paper we primarily consider the case of odd-prime dimensional qudits.

\subsection{Definition of the phase space}
Recall from Ref.~\cite{RaussendorfZurel2020} a few definitions.
\begin{Definition}\label{Definition:ClosedUnderInference}
	A set $\Omega\subset E$ is \emph{closed under inference} if for every pair of elements $a,b\in\Omega$ satisfying $[a,b]=0$, it holds that $a+b\in\Omega$.
\end{Definition}

The closure under inference of a set $\Omega\subset E$, denoted $\overline{\Omega}$, is the smallest subset of $E$ which is closed under inference and contains $\Omega$. When using the language of the symplectic vector space $E$, we also refer to the closure under inference of a set $\Omega$ as the orthogonal closure. This is related to what is called a partial closure in Ref.~\cite{KimAbramsky2023}.

\begin{Definition}\label{Definition:Noncontextuality}
	A \emph{noncontextual value assignment} for a set $\Omega\subset E$ is a function $\gamma:\Omega\to\mathbb{Z}_d$ that satisfies
	\begin{equation}\label{eq:noncontextuality condition}
		\gamma(a)+\gamma(b)-\gamma(a+b)=-\beta(a,b)
	\end{equation}
	for all pairs $a,b\in\Omega$ such that $[a,b]=0$. A set $\Omega\subset E$ is called \emph{noncontextual} if there exists a noncontextual value assignment for the set.\footnote{For odd $d$, $\beta\equiv0$, so the noncontextuality condition simplifies to $\gamma(a)+\gamma(b)=\gamma(a+b)$ for all $a,b$ such that $[a,b]=0$.}
\end{Definition}

As a result of the Mermin square proof of contextuality~\cite{Mermin1993}, the set $E$ of all Pauli observables does not satisfy Definition~\ref{Definition:Noncontextuality} when the qudit Hilbert space dimension is $d=2$ and the number of qudits is $n\ge2$. Similar proofs of contextuality can be constructed using the multiqudit Pauli observables on qudits of any even dimension, but no such proofs exist for Pauli observables on odd-dimensional qudits~\cite{OkayRaussendorf2017,RaussendorfFeldmann2023}.

A set which is both closed under inference and noncontextual we call CNC for short. The CNC phase space $\mathcal{V}$ consists of all pairs $(\Omega,\gamma)$ where $\Omega\subset E$ is a CNC set and $\gamma:\Omega\to\mathbb{Z}_d$ is a noncontextual value assignment. For any CNC set $\Omega$ and any noncontextual value assignment $\gamma:\Omega\to\mathbb{Z}_d$, we define a phase space point operator
\begin{equation}\label{eq:CNCPhasePointOperator}
	A_\Omega^\gamma=\frac{1}{d^n}\sum\limits_{b\in\Omega}\omega^{-\gamma(b)}T_b.
\end{equation}

These operators satisfy \hyperlink{QR1}{(QR1)}--\hyperlink{QR3}{(QR3)}, i.e., they define a quasiprobability function where the representation of states is given by the coefficients in the expansion
\begin{equation}\label{Equation:CNCStates}
	\rho=\sum\limits_{(\Omega,\gamma)\in\mathcal{V}}W_\rho(\Omega,\gamma)A_\Omega^\gamma.
\end{equation}
In Ref.~\cite{RaussendorfZurel2020} it was shown that for multiple qubits, this representation also satisfies \hyperlink{QR4}{(QR4)} and \hyperlink{QR5}{(QR5)}. For the case of odd dimensions, we return to the question of how the dynamics of QCM, Clifford gates and Pauli measurements, are represented in the CNC model in Section~\ref{Section:ClassicalSimulation}.

\subsection{Extremal phase space points}

By including the CNC set $\Omega$ over which the noncontextual value assignments are defined as an extra varying parameter, the phase space becomes much larger than one would expect for a Wigner function. As a result, the representation of states in Eq.~\eqref{Equation:CNCStates} is not unique. For the case of multiple qubits, when all pairs $(\Omega,\gamma)$ are included, the phase space contains redundancy. It is convenient to reduce the size of the phase space by eliminating this redundancy. This is achieved by the following lemma.
\begin{Lemma}\label{Lemma:PRARedundancyQubits}
	A CNC set $\Omega\subset E$ is called maximal if it is not strictly contained in a larger CNC set. Let $\mathcal{V}_M$ denote the multiqubit phase space consisting of pairs $(\Omega,\gamma)$ where $\Omega$ is a maximal CNC set, and $\gamma$ is a noncontextual value assignment on $\Omega$, satisfying Definitions~\ref{Definition:ClosedUnderInference} and~\ref{Definition:Noncontextuality}.  Then for any $(\tilde{\Omega},\tilde{\gamma})\in\mathcal{V}$ where $\tilde{\Omega}$ is not maximal, there are nonnegative coefficients $c(\Omega,\gamma)\ge0\;\forall(\Omega,\gamma)\in\mathcal{V}_M$ such that
	\begin{equation}\label{Equation:MaximalCNCLemma}
		A_{\tilde{\Omega}}^{\tilde{\gamma}}=\sum\limits_{(\Omega,\gamma)\in\mathcal{V}_M}c(\Omega,\gamma)A_\Omega^\gamma.
	\end{equation}
	Further, a multiqubit state $\rho$ is positively representable with respect to $\mathcal{V}$ if and only if it is positively representable with respect to $\mathcal{V}_M$.
\end{Lemma}

The proof of this lemma was given first in Ref.~\cite[\S3.3]{Zurel2020}. We include it here for completeness.

\smallskip

\emph{Proof of Lemma~\ref{Lemma:PRARedundancyQubits}.} For any CNC set $\tilde{\Omega}$ and value assignment $\tilde{\gamma}$ on $\tilde{\Omega}$, we have a phase space point operator $A_{\tilde{\Omega}}^{\tilde{\gamma}}$.  If $\tilde{\Omega}$ is not a maximal set, then as a result of the classification of \cite[Theorem 1]{RaussendorfZurel2020}, $\tilde{\Omega}$ has the form
\begin{equation}
	\tilde{\Omega}=\bigcup\limits_{k=1}^\zeta\langle a_k,I\rangle
\end{equation}
where $I$ is an isotropic subspace of $E$, all $a_k$ commute with all elements of $I$, and all $a_k$ pair-wise anti-commute. Also, $\tilde{\Omega}$ is contained in a maximal CNC set
\begin{equation}
	\Omega=\bigcup\limits_{k=1}^\xi\langle a_k,I\rangle
\end{equation}
with the same commutation structure and with $\xi>\zeta$.  Define two value assignments $\gamma_0$ and $\gamma_1$ on $\Omega$ as follows: $\gamma_0(b)=\gamma_1(b)=\gamma(b)$ for each $b\in\tilde{\Omega}$, and for each $b\in\{a_{\zeta+1},\dots,a_{\xi}\}$, define $\gamma_0(b)=0$ and $\gamma_1(b)=1$.  The values of $\gamma_0$ and $\gamma_1$ on the remaining elements of $\Omega\setminus\tilde{\Omega}$ are determined by Eq.~\eqref{eq:noncontextuality condition}.  Then,
\begin{align}
	\frac{1}{2}\left(A_\Omega^{\gamma_0}+A_\Omega^{\gamma_1}\right)&=\frac{1}{2^{n+1}}\sum\limits_{b\in\Omega}\left((-1)^{\gamma_0(b)}+(-1)^{\gamma_1(b)}\right)T_b\\
	&=\frac{1}{2^n}\sum\limits_{b\in\tilde{\Omega}}(-1)^{\gamma(b)}T_b+\frac{1}{2^{n+1}}\sum\limits_{b\in\Omega\setminus\tilde{\Omega}}\left((-1)^{\gamma_0(b)}+(-1)^{\gamma_1(b)}\right)T_b
\end{align}

For each $b\in\Omega\setminus\tilde{\Omega}$, we have one of two cases:\\
\emph{Case 1:} If $b\in\{a_{\zeta+1},\dots a_\xi\}$ then by definition $\gamma_0(b)=0$ and $\gamma_1(b)=1$.\\
\emph{Case 2:} If $b\not\in\{a_{\zeta+1},\dots a_\xi\}$ then $b=a_j+g$ for some $j\in\{\zeta+1,\dots,\xi\}$ and $g\in I$.  Then, with all addition mod $2$, we have
\begin{equation}
	\gamma_0(b)=\beta(a_j,g)+\gamma_0(a_j)+\gamma(g)=\beta(a_j,g)+\gamma(g)
\end{equation}
and
\begin{equation}
	\gamma_1(b)=\beta(a_j,g)+\gamma_1(a_j)+\gamma(g)=\beta(a_j,g)+1+\gamma(g).
\end{equation}
In both cases, $\gamma_1(b)\equiv\gamma_0(b)+1\mod2$.  Therefore, each term in the second sum in the expression above vanishes and we have
\begin{equation}
	\frac{1}{2}\left(A_\Omega^{\gamma_0}+A_\Omega^{\gamma_1}\right)=\frac{1}{2^n}\sum\limits_{b\in\Omega}(-1)^{\gamma(b)}T_b=A_{\tilde{\Omega}}^{\tilde{\gamma}}.
\end{equation}

If a state has a nonnegative representation with respect to $\mathcal{V}_M$, then clearly it has a nonnegative representation with respect to $\mathcal{V}$ since $\mathcal{V}_M\subseteq\mathcal{V}$. Conversely, if a state $\rho$ has a nonnegative expansion with respect to $\mathcal{V}$ with a positive coefficient on $A_{\tilde{\Omega}}^{\tilde{\gamma}}$, substituting the right hand side of Eq.~\eqref{Equation:MaximalCNCLemma} for $A_{\tilde{\Omega}}^{\tilde{\gamma}}$ does not introduce any negativity. Therefore, if a state is positively representable with respect to $\mathcal{V}$ then it is positively representable with respect to $\mathcal{V}_M$. $\Box$

This lemma relies on the fact that any noncontextual value assignment on $\tilde{\Omega}$ can be extended to a noncontextual value assignment on the larger CNC set $\Omega$. This is true for qubits, but in general this extension is not possible for higher dimensional qudits. See Ref.~\cite{RaussendorfZurel2020Erratum} for an example of where this fails.

The only maximal CNC set for odd-dimensional qudits is the full set of Pauli observables $E$. If we were to restrict the phase space to include only maximal CNC sets, for $n\ge2$ we obtain exactly the phase space of the Wigner function~\cite{DelfosseRaussendorf2017}. But since Lemma~\ref{Lemma:PRARedundancyQubits} fails for odd-dimensional qudits, we cannot restrict to maximal CNC sets without losing expressibility in terms of the quantum states that are postitively representable. In the next section we characterize all CNC phase space points for odd-prime-dimensional qudits. The phase space includes the phase space points of the Wigner function, but it also includes pairs $(\Omega,\gamma)$ where $\Omega\subsetneq E$ where $\gamma$ cannot be extended to a noncontextual value assignment on $E$.

\section{Characterization of phase space points}\label{Section:PhaseSpaceCharacterization}

Our goal in this section is to characterize the CNC phase space for odd-prime-dimensional qudits, as was achieved for the case of qubits in Ref.~\cite[\S IV]{RaussendorfZurel2020}. This requires characterizing all pairs $(\Omega,\gamma)$, where $\Omega\subset E$ is a CNC set and $\gamma:\Omega\to\mathbb{Z}_d$ is a noncontextual value assignment for $\Omega$, since points in the phase space are identified with these pairs. To start, we characterize the CNC sets.

\subsection{Characterization of CNC sets}

Note that there are no state-independent proofs of contextuality like the Mermin square~\cite{Mermin1993} using only odd-dimensional Pauli observables~\cite{OkayRaussendorf2017,SchmidPusey2022,RaussendorfFeldmann2023}. Therefore, for the purpose of characterizing CNC sets, we can ignore the noncontextuality condition of Definition~\ref{Definition:Noncontextuality}, and focus only on the closure under inference condition, Definition~\ref{Definition:ClosedUnderInference}. We now characterize the sets $\Omega\subset E$ that are closed under inference. A similar characterization for the case $d=2$ has already been found~\cite{Heimendahl2019,KirbyLove2019,RaussendorfZurel2020}. For the case of odd prime $d$, this is achieved by the following theorem.
\begin{Theorem}\label{theorem:classification-cnc-qudits}
	For any number of qudits $n$ of any odd prime dimension $d$, a set $\Omega\subset E$ is closed under inference if and only if
	\begin{enumerate}
		\item [(i)] $\Omega$ is a subspace of $E$ (i.e. $\Omega$ is closed under the addition), or
		\item [(ii)] $\Omega$ has the form 
		\begin{equation}\label{eq:shape-of-cnc-sets}
			\Omega=\bigcup\limits_{k=1}^{\xi}\Span{a_k,I}
		\end{equation}
		where $I\subset E$ is an isotropic subspace and the generators $a_1,\dots,a_\xi$ satisfy $[a_i,a_j]\neq0$ for all $i\neq j$, and $a_i\in I^\perp$ for all $i\in\{1,\dots,\xi\}$.
	\end{enumerate}
\end{Theorem}

The proof of this theorem requires a couple of lemmas.

\begin{Lemma}\label{lemma:Adding-Element-to-subspace}
	Let $\Omega\subset E$ be a subspace and $v\in E$. Then $\overline{\Omega\cup\{v\}}$ has the form Eq.~\eqref{eq:shape-of-cnc-sets} if and only if $\Omega\cap v^\perp$ is isotropic. Otherwise, $\overline{\Omega\cup\{v\}}=\Span{v,\Omega}$.
\end{Lemma}

\begin{Lemma}\label{lemma:Adding-element-to-cnc-set}
	Suppose $\Omega\subset E$ has the form Eq.~\eqref{eq:shape-of-cnc-sets}, and $v\in E$. Then $\overline{\Omega\cup\{v\}}$ has the form Eq.~\eqref{eq:shape-of-cnc-sets}, or it is a subspace. 
\end{Lemma}

The proofs of these lemmas are in Appendix~\ref{Section:Proofs}.

\smallskip

\emph{Proof of Theorem~\ref{theorem:classification-cnc-qudits}.}
It is easy to verify that subspaces and sets of the form Eq.~\eqref{eq:shape-of-cnc-sets} are closed under inference. The proof for the $d=2$ case is given by Lemma~3 of Ref.~\cite{RaussendorfZurel2020}). The proof is identical for the case of odd prime $d$. Here we focus on the other direction, that every set that is closed under inference is either a subspace or has the form described by Eq.~\eqref{eq:shape-of-cnc-sets}.

The overall proof strategy is induction. For the base case, start with one element $u\in E$. Obviously $\overline{\{u\}}=\langle u\rangle$, which is simultaneously a subspace and of the form Eq.~\eqref{eq:shape-of-cnc-sets}. In the induction step, suppose that a set $\Omega\subset E$ is a subspace or of the form \eqref{eq:shape-of-cnc-sets}. Then we show that for all $v\in E$ the orthogonal closure $\overline{\Omega \cup \{v \}}$ is again a subspace or of the form \eqref{eq:shape-of-cnc-sets}. The induction step follows immediately from Lemmas~\ref{lemma:Adding-Element-to-subspace} and~\ref{lemma:Adding-element-to-cnc-set}. $\Box$

\subsection{Maximal size of cnc sets}

For a set $\Omega$ of the form Eq.~\eqref{eq:shape-of-cnc-sets}, once the isotropic subspace $I$ is fixed, the elements $a_1,\dots,a_\xi$ are chosen from $I^\perp/I\simeq\mathbb{Z}_d^{2(n-\dim(I))}$. Then the size of the set is $(\xi d-\xi+1)\cdot d^{\dim(I)}$. That is, it depends on the value of $\xi$, which is bounded by the maximal number of mutually non-orthogonal elements in $\mathbb{Z}_d^{2(n-\dim(I))}$. The maximal number of mutually non-orthogonal elements in $\mathbb{Z}_d^{2n}$ is unknown~\cite{Chin2005}, but we can prove that it is at least $dn+1$.\footnote{In order to prove efficiency of the classical simulation algorithm in Section~\ref{Section:ClassicalSimulation}, we need the number $\xi$ of the noncommuting generators of the CNC sets to be polynomial in the number $n$ of qudits. We expect the maximal number of pair-wise noncommuting elements in $\mathbb{Z}_d^{2n}$ to be bounded by a polynomial in $n$, but as noted this has not been proven. If it turns out that this conjecture is false, then in order for the algorithm to be efficient we must explicitly impose a polynomial bound on the value $\xi$ for the admissible CNC phase space points used in the simulation. See Section~\ref{Section:ClassicalSimulation} for details.} This can be seen as a generalization of the corresponding qubit construction in~\cite[Theorem~1]{RaussendorfZurel2020}. 

Let $e_1,\dots,e_n,f_1,\dots f_n$ be the standard basis of the symplectic vector space $E_n\simeq\mathbb{Z}_d^{2n}$ ($[e_i,e_j]=0$, $[f_i,f_j]=0$, and $ [e_i,f_j]=\delta_{i,j}$). The space $E_1$ can be decomposed into $d+1$ mutually non-orthogonal lines as
\begin{equation}\label{eq:Covering-of-E_1-by-mutually-non-orthogonal-lines}
	E_1=\Span{c_1}\cup\cdots\cup\Span{c_{d+1}},\quad[c_i,c_j]\neq0\text{ for all }i\neq j.
\end{equation}
For example, such a decomposition if given by 
\begin{equation}
	E_1=\Span{e_1}\cup\Span{f_1}\cup\Span{e_1+f_1}\cup\Span{e_1+2f_1}\cup\cdots\cup\Span{e_1+(d-1)f_1}.
\end{equation}
This gives rise to the following construction for any $n\in\mathbb{N}$ (which is in a sense a generalization of the construction given in \cite[eqs.~(16),~(17)]{RaussendorfZurel2020}). Labeling the elements of $E_n$ by $(u_1,\ldots,u_n)$ with $ u_i\in E_1$ for $i = 1,\ldots,n$. We have the following construction of $dn+1$ mutually nonorthogonal lines,
\begin{equation}\label{eq:GeneralizationOfMajorana}
	\begin{cases}
		\begin{array}{llll}
			(c_1, 0,\dots,0), & (c_2,0, \dots,0), & \dots, & (c_d,0, \dots,0), \\
			(c_{d+1},c_1,0\dots,0), & (c_{d+1},c_2,0,\dots,0), & \dots, & (c_{d+1},c_d,0,\dots,0), \\
			(c_{d+1},c_{d+1},c_1,0\dots,0), & (c_{d+1},c_{d+1},c_2,0,\dots,0), & \dots, & (c_{d+1},c_{d+1},c_d,0,\dots,0), \\
			\vdots&&&\\
			(c_{d+1},c_{d+1}\dots,c_{d+1},c_1), & (c_{d+1},c_{d+1}\dots,c_{d+1},c_2), & \dots, & (c_{d+1},c_{d+1}\dots,c_{d+1},c_d), \\
			(c_{d+1},c_{d+1}\dots,c_{d+1},c_{d+1}). &&&
		\end{array}
	\end{cases}
\end{equation}

It is straightforward to check that all of the given generators are mutually non-orthogonal. The symplectic inner product of two generators in the same row above is proportional to $[c_i,c_j]\neq0$ for all $i\neq j$. The symplectic inner product of two elements that lie in different rows above will be proportional to $[c_i,c_{d+1}]\neq0$ for some $i\in\{1,\dots,d\}$.

\subsection{Characterization of noncontextual value assignments on CNC sets}

Theorem~\ref{theorem:classification-cnc-qudits} gives a characterization of all CNC sets $\Omega$. With a characterization of the noncontextual value assignments $\gamma$ on $\Omega$, we obtain a characterization of the phase space $\mathcal{V}$. The proof of \cite[Lemma~2]{RaussendorfZurel2020} can be adapted to show that the value assignments are a coset of a vector space, and in fact, since $\beta\equiv0$ in odd dimensions, they are a vector space. This characterization can be made more explicit.

First, for the case $\Omega=E$ with $n\ge2$, Ref.~\cite[Lemma~1]{DelfosseRaussendorf2017} gives a characterization of the value assignments. The value assignments $\gamma:\Omega\to\Z_d$ are linear functions (equivalently, $ \omega^{\gamma(\cdot)} $ are characters of $\Omega$). When $\Omega$ is a subspace of $E$, the restriction of these characters to $\Omega$ are still value assignments. The single-qudit case is an exception, with more value assignments than characters (see \cite[\S3.4]{DelfosseRaussendorf2017}). Second, in the case $\Omega$ has the form Eq.~\eqref{eq:shape-of-cnc-sets} with $\xi\ge2$, we can describe the value assignments as follows. The value assignment $\gamma$ can be chosen freely on $a_1,\dots,a_\xi$, as well as on a basis for $I$. Then the value of $\gamma$ on the rest of $\Omega$ is determined through Eq.~\eqref{eq:noncontextuality condition}.

If a value assignment on $\Omega$ happens to be linear, then it can be extended by linearity to a value assignment on $E$ of all Pauli operators. In this case, Lemma~\ref{Lemma:PRARedundancyQubits} can be adapted to show that the corresponding phase space point $(\Omega,\gamma)$ is redundant (the proof given in \cite[Lemma~1]{RaussendorfZurel2020} holds). But if $\gamma$ is not linear on $\Omega$, then it cannot be extended to a value assignment on $E$, and the phase space point $(\Omega,\gamma)$ is not necessarily redundant. See Ref.~\cite{RaussendorfZurel2020Erratum} for an example of a phase space point $(\Omega,\gamma)$ with a nonlinear value assignment $\gamma$. Therefore, there exist phase space points in the CNC construction that do not correspond to phase space points of Gross' Wigner function or convex mixtures thereof, namely, those with $\Omega\subsetneq E$ and nonlinear value assignments $\gamma:\Omega\to\mathbb{Z}_d$.

\section{Extended classical simulation algorithm}\label{Section:ClassicalSimulation}

In this section we show that the CNC construction provides a classical simulation algorithm for quantum computation with magic states. To start we have show how CNC-type phase space point operators are updated under Clifford gates and Pauli measurements, and from these update rules, the classical simulation algorithm follows. It is given explicitly in Algorithm~\ref{Algorithm:simAlg}. Furthermore, with the characterization of the phase space given in Section~\ref{Section:PhaseSpaceCharacterization}, we are able to show that the update of phase space points under Clifford gates and Pauli measurements can be computed efficiently classically. This is the result of Theorem~\ref{Theorem:UpdateRuleEfficiency}.

\pagebreak

\subsection{State update under Clifford gates}\label{Section:CliffordUpdate}

Using the action of the Clifford group on the Pauli operators in Eq.~\eqref{Equation:CliffordActionPaulis}, for every CNC set $\Omega$, each noncontextual value assignment $\gamma:\Omega\to\mathbb{Z}_d$, and every Clifford group element $g\in\Cl$, we introduce the derived objects
\begin{equation}
	g\cdot\Omega:=\{S_ga\;|\;a\in\Omega\}
\end{equation}
and $g\cdot\gamma:g\cdot\Omega\to\mathbb{Z}_d$ defined by
\begin{equation}
	g\cdot\gamma(S_gb)=\gamma(b)-\Phi_g(b),\quad\forall b\in\Omega.
\end{equation}

\begin{Lemma}\label{Lemma:CliffordUpdate}
	If $\Omega\subset E$ is CNC and $\gamma:\Omega\to\mathbb{Z}_d$ is a noncontextual value assignment for $\Omega$, then for any Clifford group element $g\in\Cl$,
	\begin{equation}
		gA_\Omega^\gamma g^\dagger=A_{g\cdot\Omega}^{g\cdot\gamma},
	\end{equation}
	where $g\cdot\Omega$ is CNC and $g\cdot\gamma$ is a noncontextual value assignment for $g\cdot\Omega$.
\end{Lemma}

{\emph{Proof of Lemma~\ref{Lemma:CliffordUpdate}.} For any $a,b\in g\cdot\Omega$ with $[a,b]=0$, there exist $c,d\in\Omega$ such that $a=S_gc$ and $b=S_gd$, and since $S_g\in\Sp(E)$, $[c,d]=0$.  Since $\Omega$ is closed under inference, $c+d\in\Omega$, therefore, $S_g(c+d)=a+b\in g\cdot\Omega$.  Thus, $g\cdot\Omega$ is closed under inference. $g\cdot\gamma$ satisfies Eq.~\eqref{eq:noncontextuality condition} by Lemma~4 in Ref.~\cite{OkayRaussendorf2017}. Therefore, $g\cdot\gamma:g\cdot\Omega\to\mathbb{Z}_d$ is a noncontextual value assignment for $g\cdot\Omega$.

Finally, we calculate
\begin{align}
	gA_\Omega^\gamma g^\dagger=&\frac{1}{d^n}\sum\limits_{b\in\Omega}\omega^{-\gamma(b)}gT_bg^\dagger\\
	=&\frac{1}{d^n}\sum\limits_{b\in\Omega}\omega^{-\gamma(b)+\Phi_g(b)}T_{S_gb}\\
	=&\frac{1}{d^n}\sum\limits_{b\in g\cdot\Omega}\omega^{-g\cdot\gamma(b)}T_b=A_{g\cdot\Omega}^{g\cdot\gamma}
\end{align}
and we obtain the final statement of the lemma. $\Box$

\medskip

This lemma shows that the CNC phase space is closed under the action of the Clifford group. As a result, the corresponding quasiprobability function is Clifford covariant.

\begin{Corollary}\label{Corollary:CliffordCovariance}
	For any quantum state $\rho$ and any Clifford group element $g\in\Cl$, there exists a representation of the state $g\rho g^\dagger$ satisfying $W_{g\rho g^\dagger}(g\cdot\Omega,g\cdot\gamma)=W_\rho(\Omega,\gamma)$ for all $(\Omega,\gamma)\in\mathcal{V}$.
\end{Corollary}

The proof of this corollary follows immediately by conjugating the expression Eq.~\eqref{Equation:CNCStates} by a Clifford group element $g\in\Cl$. This also implies that the negativity in the representation does not increase under Cliffords.

\begin{Corollary}
	Suppose a state $\rho$ has an expansion as in Eq.~\eqref{Equation:CNCStates} such that $W_\rho(\Omega,\gamma)\ge0$ for all $(\Omega,\gamma)\in\mathcal{V}$. Then for any Clifford group element $g\in\Cl$, there exists an expansion of $g\rho g^\dagger$ such that $W_{g\rho g^\dagger}(\Omega,\gamma)\ge0$ for all $(\Omega,\gamma)\in\mathcal{V}$.
\end{Corollary}

\subsection{State update under Pauli measurements}\label{Section:PauliUpdate}

\begin{Lemma}\label{Lemma:PauliUpdate}
	Let $\Omega\subset E$ be a CNC set and $\gamma:\Omega\to\mathbb{Z}_d$ be a noncontextual value assignement for $\Omega$.  Then for any isotropic subspace $I\subset E$ with any noncontextual value assignment $r:I\to\mathbb{Z}_d$, either
	\begin{itemize}
		\item[(I)]$\Tr(\Pi_I^rA_\Omega^\gamma)=0$ and $\Pi_I^rA_\Omega^\gamma\Pi_I^r=0$, or
		\item[(II)]$\Tr(\Pi_I^rA_\Omega^\gamma)>0$ and
		\begin{equation}
			\frac{\Pi_I^rA_\Omega^\gamma\Pi_I^r}{\Tr(\Pi_I^rA_\Omega^\gamma)}=A_{I+\Omega\cap I^\perp}^{\gamma\times r}
		\end{equation}
		where $I+\Omega\cap I^\perp$ is CNC and $\gamma\times r:I+\Omega\cap I^\perp\to\mathbb{Z}_d$ is the unique noncontextual value assignment for $I+\Omega\cap I^\perp$ satisfying $\gamma\times r|_{I}=r$ and $\gamma\times r|_{\Omega\cap I^\perp}=\gamma|_{\Omega\cap I^\perp}$.
	\end{itemize}
\end{Lemma}

\emph{Proof of Lemma~\ref{Lemma:PauliUpdate}.} The proof is similar to the third part of the proof of Theorem~1 in Ref.~\cite{ZurelHeimendahl2024} and to the proof of Lemma~7 in Ref.~\cite{ZurelHeimendahl2024}, the difference being here $A_\Omega^\gamma$ is not in general a stabilizer code projector.  We have
\begin{align}
	\Tr(\Pi_I^rA_\Omega^\gamma)=&\frac{1}{d^n|I|}\sum\limits_{a\in I}\sum\limits_{b\in\Omega}\omega^{-r(a)-\gamma(b)}\Tr(T_aT_b)\\
	=&\frac{1}{d^n|I|}\sum\limits_{a\in I}\sum\limits_{b\in\Omega\cap I^\perp}\omega^{-r(a)-\gamma(b)}\Tr(T_aT_b) + \sum\limits_{a\in I}\sum\limits_{b\in\Omega\setminus(\Omega\cap I^\perp)}\omega^{-r(a)-\gamma(b)}\Tr(T_aT_b)
\end{align}
The second sum vanishes because $I$ is isotropic so $I$ and $\Omega\setminus(\Omega\cap I^\perp)$ are disjoint, therefore, $\Tr(T_aT_b)=0$ for all $a\in I$, $b\in\Omega\setminus(\Omega\cap I^\perp)$.  Continuing with the first sum we have
\begin{align}
	\Tr(\Pi_I^rA_\Omega^\gamma)=&\frac{1}{d^n|I|}\sum\limits_{a\in I}\sum\limits_{b\in\Omega\cap I^\perp}\omega^{-r(a)-\gamma(b)}\Tr(T_aT_b)\\
	=&\frac{1}{|I|}\sum\limits_{a\in\Omega\cap I}\omega^{-r(a)-\gamma(-a)}\\
	=&\frac{1}{|I|}\sum\limits_{a\in\Omega\cap I}\omega^{-r(a)+\gamma(a)}.
\end{align}
Here we have two cases.  First, if $r|_{\Omega\cap I}\ne\gamma|_{\Omega\cap I}$, then by orthogonality of twisted characters~\cite{Cheng2015}, we have $\Tr(\Pi_I^rA_\Omega^\gamma)=0$.  Second, if $r|_{\Omega\cap I}=\gamma|_{\Omega\cap I}$ then
\begin{align}
	\Tr(\Pi_I^rA_\Omega^\gamma)=\frac{1}{|I|}\sum\limits_{a\in\Omega\cap I}\omega^{-r(a)+\gamma(a)}=\frac{|\Omega\cap I|}{|I|}>0.
\end{align}
Therefore, either $\Tr(\Pi_I^rA_\Omega^\gamma)=0$, or $\Tr(\Pi_I^rA_\Omega^\gamma)>0$.\footnote{This implies the operators $A_\Omega^\gamma$ are contained in the $\Lambda$ polytopes of Refs.~\cite{ZurelRaussendorf2020,ZurelHeimendahl2024}.}  Here we have two cases.

(I)~In the first case, we must show that $\Pi_I^rA_\Omega^\gamma\Pi_I^r=0$.  For any Pauli operator $T_a,a\in E$, consider the inner product $\Tr(T_a\Pi_I^rA_\Omega^\gamma\Pi_I^r)$.  Here we have three subcases. (i)~First, if $a\in I$ then
\begin{align}
	T_a\Pi_I^r=&\frac{1}{|I|}\sum\limits_{b\in I}\omega^{-r(b)}T_aT_b\\
	=&\frac{1}{|I|}\sum\limits_{b\in I}\omega^{-r(b)-\beta(a,b)}T_{a+b}\\
	=&\frac{1}{|I|}\sum\limits_{b\in I}\omega^{r(a)-r(a+b)}T_{a+b}\\
	=&\omega^{r(a)}\Pi_I^r.
\end{align}
Therefore,
\begin{equation}
	\Tr(T_a\Pi_I^rA_\Omega^\gamma\Pi_I^r)=\omega^{r(a)}\Tr(\Pi_I^rA_\Omega^\gamma)=0.
\end{equation}
(ii)~Second, if $a\in I^\perp\setminus I$, then by Lemma~5 of Ref.~\cite{ZurelHeimendahl2024}
\begin{equation}
	\Pi_I^r=\sum\limits_{r'\in\Gamma_{I,r}^{\langle I,a\rangle}}\Pi_{\langle I,a\rangle}^{r'}
\end{equation}
where $\Gamma_{I,r}^{\langle I,a\rangle}$ is the set of noncontextual value assignments on $\langle I,a\rangle$ satisfying $r'|_I=r$.  Multiplying this equation by $A_\Omega^\gamma$ and taking a trace we get
\begin{equation}
	\Tr(\Pi_I^rA_\Omega^\gamma)=\sum\limits_{r'\in\Gamma_{I,r}^{\langle I,a\rangle}}\Tr(\Pi_{\langle a,I\rangle}^{r'}A_\Omega^\gamma).
\end{equation}
The left hand side is zero by assumption.  As shown above, each term on the right hand side is nonnegative, therefore each term on the right hand side is zero, i.e. $\Tr(\Pi_{\langle a,I\rangle}^{r'}A_\Omega^\gamma)=0$ for all $r'\in\Gamma_{I,r}^{\langle a,I\rangle}$.

We can write the spectral decomposition of $T_a$ as
\begin{equation}
	T_a=\sum\limits_{r'\in\Gamma_{I,r}^{\langle a,I\rangle}}\omega^{r(a)}\Pi_{\langle a\rangle}^{r'|_{\langle a\rangle}}.
\end{equation}
Then
\begin{align}
	\Tr(T_a\Pi_I^rA_\Omega^\gamma\Pi_I^r)=\sum\limits_{r'\in\Gamma_{I,r}^{\langle a,I\rangle}}\omega^{r'(a)}\Tr(\Pi_{\langle a\rangle}^{r'|_{\langle a\rangle}}\Pi_I^rA_\Omega^\gamma\Pi_I^r)=\sum\limits_{r'\in\Gamma_{I,r}}\omega^{r'(a)}\Tr(\Pi_{\langle a,I\rangle}^{r'}A_\Omega^\gamma)=0.
\end{align}

(iii)~Third, if $a\notin I^\perp$ then
\begin{align}
	\Pi_I^rT_a\Pi_I^r=&\frac{1}{|I|^2}\sum\limits_{b,c\in I}\omega^{-r(b)-r(c)}T_bT_aT_c\\
	=&\frac{1}{|I|^2}\sum\limits_{b,c\in I}\omega^{-r(b)-r(c)+[b,a]-\beta(b,c)}T_aT_{b+c}\\
	=&\frac{1}{|I|^2}T_a\sum\limits_{b,c\in I}\omega^{-r(b+c)+[b,a]}T_{b+c}\\
	=&\frac{1}{|I|}T_a\Pi_I^r\sum\limits_{b\in I}\omega^{[b,a]}.
\end{align}
By character orthogonality, the sum in the final expression vanishes.  Therefore,
\begin{equation}
	\Tr(T_a\Pi_I^rA_\Omega^\gamma\Pi_I^r)=\Tr(\Pi_I^rT_a\Pi_I^rA_\Omega^\gamma)=0.
\end{equation}
Thus, $\Tr(T_a\Pi_I^rA_\Omega^\gamma\Pi_I^r)=0$ for all Pauli operators $T_a,a\in E$, and so $\Pi_I^rA_\Omega^\gamma\Pi_I^r=0$.

(II)~In the second case, $\Tr(\Pi_I^rA_\Omega^\gamma)>0$, as shown above we have $r|_{\Omega\cap I}=\gamma|_{\Omega\cap I}$.  Then
\begin{align}
	\Pi_I^rA_\Omega^\gamma\Pi_I^r=&\frac{1}{|I|d^n}\sum\limits_{b\in\Omega}\sum\limits_{c\in I}\omega^{-\gamma(b)-r(c)}\Pi_I^rT_bT_c\\
	=&\frac{1}{|I|d^n}\sum\limits_{b\in\Omega}\sum\limits_{c\in I}\omega^{-\gamma(b)-r(c)+[b,c]}\Pi_I^rT_cT_b\\
	=&\frac{1}{|I|d^n}\sum\limits_{b\in\Omega}\sum\limits_{c\in I}\omega^{-\gamma(b)+[b,c]}\Pi_I^rT_b\\
	=&\frac{1}{|I|d^n}\Pi_I^r\sum\limits_{b\in\Omega}\left[\sum\limits_{c\in I}\omega^{[b,c]}\right]\omega^{-\gamma(b)}T_b\\
	=&\frac{1}{d^n}\Pi_I^r\sum\limits_{b\in\Omega\cap I^\perp}\omega^{-\gamma(b)}T_b\\
	=&\frac{1}{|I|d^n}\sum\limits_{a\in I}\sum\limits_{b\in\Omega\cap I^\perp}\omega^{-r(a)-\gamma(b)}T_aT_b\\
	=&\frac{1}{|I|d^n}\sum\limits_{a\in I}\sum\limits_{b\in\Omega\cap I^\perp}\omega^{-r(a)-\gamma(b)-\beta(a,b)}T_{a+b}
\end{align}
Here each Pauli operator in the sum has the same multiplicity.  For any $c\in I+\Omega\cap I^\perp$, let $\mu(c)$ denote the number of pairs $(a,b)\in I\times \Omega\cap I^\perp$ such that $a+b=c$.  We will show that $\mu(c)=\mu(0)$ for all $c\in I+\Omega\cap I^\perp$.  First, suppose $(a_1,b_1),(a_2,b_2),\dots,(a_{\mu(c)},b_{\mu(c)})$ are $\mu(c)$ distinct pairs in $I\times\Omega\cap I^\perp$ such that $a_j+b_j=c$.  Then the pairs $(a_j-a_1,b_j-b_1)\in I\times\Omega\cap I^\perp$ for $j=2,3,\dots,\mu(c)$, together with the pair $(0,0)$ show that $\mu(0)\ge\mu(c)$ for any $c\in I+\Omega\cap I^\perp$. Now let $(a_1,b_1),(a_2,b_2),\dots,(a_{\mu(0)},b_{\mu(0)})$ be distinct pairs in $I\times\Omega\cap I^\perp$ such that $a_j+b_j=0$, and let $(a,b)$ be such that $a+b=c$.  Then the pairs $(a_j+a,b_j+b)$ for $j=1,2,\dots,\mu(0)$ show that $\mu(c)\ge\mu(0)$ for any $c\in I+\Omega\cap I^\perp$. Thus, $\mu(c)=\mu(0)$ for any $c\in I+\Omega\cap I^\perp$.  We have $\mu(0)=|\Omega\cap I|$.

The noncontextual value assignment $\gamma\times r:I+\Omega\cap I^\perp\to\mathbb{Z}_d$ satisfying $\gamma\times r|_I=r$ and $\gamma\times r|_{\Omega\cap I^\perp}=\gamma|_{\Omega\cap I^\perp}$ is uniquely defined.  Then
\begin{align}
	\Pi_I^rA_\Omega^\gamma\Pi_I^r=\frac{|\Omega\cap I|}{|I|d^n}\sum\limits_{a\in I+\Omega\cap I^\perp}\omega^{-\gamma\times r(a)}T_a=\frac{|\Omega\cap I|}{|I|}A_{I+\Omega\cap I^\perp}^{\gamma\times r}
\end{align}
Thus
\begin{equation}
	\frac{\Pi_I^rA_\Omega^\gamma\Pi_I^r}{\Tr(\Pi_I^rA_\Omega^\gamma)}=A_{I+\Omega\cap I^\perp}^{\gamma\times r}
\end{equation}
which concludes the proof of the lemma. $\Box$

For a single Pauli measurement, $I=\langle a\rangle$ for some $a\in E$, this lemma simplifies with the following corollary.
\begin{Corollary}
	Let $(\Omega,\gamma)\in\mathcal{V}$ be a CNC pair. Then for any $a\in E$,
	\begin{itemize}
		\item[(I)] if $a\in\Omega$, then
			$\begin{cases}
				\Tr(\Pi_a^sA_\Omega^\gamma)=1\text{ and }\Pi_a^sA_\Omega^\gamma\Pi_a^s=A_{\Omega\cap a^\perp}^{\gamma|_{\Omega\cap a^\perp}}\quad\text{if $s=\gamma(a)$,}\\
				\Tr(\Pi_a^sA_\Omega^\gamma)=0\text{ and }\Pi_a^sA_\Omega^\gamma\Pi_a^s=0\quad\text{if $s\ne\gamma(a)$.}
			\end{cases}$
		\item[(II)] if $a\notin\Omega$, then $\Tr(\Pi_a^sA_\Omega^\gamma)=1/d$ for each $s\in\mathbb{Z}_d$, and $\Pi_a^sA_\Omega^\gamma\Pi_a^s=\frac{1}{d}A_{\langle a\rangle+\Omega\cap a^\perp}^{\gamma\times s}$.
	\end{itemize}
\end{Corollary}

\subsection{Classical simulation algorithm}

The update rules of the phase space points under Clifford gates and Pauli measurements given in Sections~\ref{Section:CliffordUpdate} and~\ref{Section:PauliUpdate} allow us to introduce a classical simulation algorithm for quantum computation with magic states. This simulation is given explicitly in Algorithm~\ref{Algorithm:simAlg}. The proof of correctness of this algorithm is analogous to the proofs of~\cite[Theorem 3]{RaussendorfZurel2020},~\cite[Theorem~2]{ZurelRaussendorf2020}, and \cite[Theorem~2]{ZurelHeimendahl2024}.\footnote{One difference between the simulation Algorithm~\ref{Algorithm:simAlg} and that of Ref.~\cite{RaussendorfZurel2020} is there only extremal (in the sense of Lemma~\ref{Lemma:PRARedundancyQubits}) CNC phase space points are used. Clifford gates preserve the property of extremality and after each Pauli measurement the resulting post-measurement state $\Pi_a^sA_\Omega^\gamma\Pi_a^s/\Tr(\Pi_a^sA_\Omega^\gamma)$ is decomposed as a convex combination of extremal CNC phase space point operators, c.f.~\cite[Equation~26]{RaussendorfZurel2020}. On the other hand, in our case we allow nonextremal CNC points to appear in the simulation, this does not affect the proof of correctness of the algorithm.}

\begin{algorithm}[H]
	\begin{algorithmic}[1]
		\REQUIRE $W_{\rho}(\Omega,\gamma)$; a Clifford+Pauli circuit $\mathcal{C}$
		\STATE sample a point $(\Omega,\gamma)\in\mathcal{V}$ according to the probability distribution $p_{\rho}$
		\WHILE{end of circuit has not been reached}
		\IF{a Clifford unitary $g\in\mathcal{C}\ell$ is encountered}
		\STATE update $(\Omega,\gamma)\leftarrowtail(g\cdot\Omega,g\cdot\gamma)$
		\ENDIF
		\IF{a Pauli measurements $T_a,\;a\in E$ is encountered}
		\IF{$a\in\Omega$}
		\STATE set $s=\gamma(a)$
		\STATE update $(\Omega,\gamma)\leftarrowtail(\Omega\cap a^\perp,\gamma|_{\Omega\cap a^\perp})$
		\ELSIF{$a\not\in\Omega$}
		\STATE choose a random $s\in\mathbb{Z}_d$ distributed uniformly
		\STATE update $(\Omega,\gamma)\leftarrowtail(\Span{a}+\Omega\cap a^\perp,\gamma\times s)$
		\ENDIF
		\STATE \textbf{Output:} $s$ as the outcome of the measurement
		\ENDIF
		\ENDWHILE
	\end{algorithmic}
	\caption{One run of the classical simulation algorithm for quantum computation with magic states. The algorithm provides samples from the joint probability distribution of the Pauli measurements in a quantum circuit consisting of Clifford gates and Pauli measurements applied to an input state $\rho$.\label{Algorithm:simAlg}}
\end{algorithm}

\subsection{Efficiency of classical simulation}\label{Section:Efficiency}

In the case of the multiqubit CNC construction, the classical simulation algorithm~\cite[Table~1]{RaussendorfZurel2020} is efficient on states with a nonnegative quasiprobability function $W_\rho$. The proof of this fact relies on the structure of the multiqubit CNC phase space points found in~\cite[\S IV]{RaussendorfZurel2020}. With the structure of the phase space found in Section~\ref{Section:PhaseSpaceCharacterization}, an adaptation of the proof of efficiency should apply to our case, with one caveat. Namely, for multiple qubits, the maximum number of pair-wise noncommuting Pauli operators on $n$ qubits is known to be $2n+1$. In particular, it is linear in the number of qubits. For odd-prime-dimensional qudits, a polynomial bound on this number is not known, the best known upper bound is exponential in the number of qudits $n$~\cite{Chin2005}. If there were to exist sets of noncommuting Pauli operators of size exponential in $n$, the classical updates of the CNC phase space points would not be efficiently computable classically.

Let $\tilde{\mathcal{V}}$ denote a subset of the CNC phase space defined by the additional constraint that the CNC sets have a set of generators (consisting of a basis in case of a subspace of $E$, or a basis of $I$ together with the noncommuting generators $a_1,\dots,a_\xi$ in the case of Eq.~\eqref{eq:shape-of-cnc-sets}, c.f. Theorem~\ref{theorem:classification-cnc-qudits}) that has size polynomial in the number of qudits. The state updates under Clifford gates and Pauli measurements described in Lemmas~\ref{Lemma:CliffordUpdate} and~\ref{Lemma:PauliUpdate} (lines~4,~9, and~12 of Algorithm~\ref{Algorithm:simAlg}) do not increase the number of generators needed to specify a CNC set, so $\tilde{\mathcal{V}}$ is closed under these operations. With respect this this phase space $\tilde{\mathcal{V}}$, we have the following theorem.

\begin{Theorem}\label{Theorem:UpdateRuleEfficiency}
	For any number of qudits $n$ of any odd prime Hilbert space dimension $d$,
	\begin{enumerate}
		\item The update of phase space points $\tilde{\mathcal{V}}$ under Clifford gates given by Lemma~\ref{Lemma:CliffordUpdate} can be computed in time polynomial in the number of qudits.
		\item The update of phase space points $\tilde{\mathcal{V}}$ under Pauli measurements given by Lemma~\ref{Lemma:PauliUpdate} can be computed in time polynomial in the number of qudits.
	\end{enumerate}
\end{Theorem}

The proof of Theorem~\ref{Theorem:UpdateRuleEfficiency} is analogous to the proof of Theorem~3 in Ref.~\cite{RaussendorfZurel2020}. Basically, with the characterization of phase space points in Section~\ref{Section:PhaseSpaceCharacterization}, for any phase space points $(\Omega,\gamma)$, the set $\Omega$ can be identified by a small number of vectors in $\mathbb{Z}_d^{2n}$. Namely, if $\Omega$ is a subspace of $E$ then we identify $\Omega$ by a basis of vectors, and if $\Omega$ has the form Eq.~\eqref{eq:shape-of-cnc-sets} then we use the vectors $a_1,\dots,a_\xi$ in Eq.~\eqref{eq:shape-of-cnc-sets} as well as a basis for $I$. Further, the value assignment $\gamma$ can be specified by specifying its value on these generating vectors, and then its value on the rest of $\Omega$ is determined through Eq.~\eqref{eq:noncontextuality condition}. Then all of the steps in the update can be performed by linear algebra in $\mathbb{Z}_d^{2n}$ on this small number of vectors.

\begin{Corollary}\label{Theorem:SimulationEfficiency}
	For any quantum circuit consisting of a sequence of polynomially many Clifford gates and Pauli measurements on an $n$-qudit input state $\rho$, satisfying the following conditions:
	\begin{itemize}
		\item $\rho=\sum\limits_{(\Omega,\gamma)\in\tilde{\mathcal{V}}}W_\rho(\Omega,\gamma)A_\Omega^\gamma$ where $W_\rho(\Omega,\gamma)\ge0$ for all $(\Omega,\gamma)\in\tilde{\mathcal{V}}$,
		\item samples from the distribution $W_\rho$ can be obtained efficiently,
	\end{itemize}
	the outcome distributions of the measurements can be efficiently sampled from using Algorithm~\ref{Algorithm:simAlg}.
\end{Corollary}

Of course, it could turn out that the maximal number of noncommuting Pauli operators on $n$ qudits is bounded by a polynomial in the number of qudits $n$, in which case Theorem~\ref{Theorem:UpdateRuleEfficiency} and Corollary~\ref{Theorem:SimulationEfficiency} apply to the full CNC phase space $\mathcal{V}$, but as mentioned above this has not been proven. This is an open problem.

\section{Relation to the \texorpdfstring{$\Lambda$}{Lambda}-polytope model}\label{Section:LambdaConnection}

The $\Lambda$ polytope models~\cite{ZurelRaussendorf2020,ZurelHeimendahl2024} are a family of quasiprobability representations for quantum computation with magic states satisfying \hyperlink{QR1}{(QR1)}--\hyperlink{QR5}{(QR5)} in which no negativity is needed in the representation of any state. That is, all elements of this model of quantum computation---all states, Clifford gates, and Pauli measurements are described probabilistically.

The qubit version of the model was extensively studied in~\cite{ZurelRaussendorf2020,OkayRaussendorf2021,Heimendahl2019}. There it was shown that for the multiqubit case ($d=2$), for maximal CNC sets $\Omega$, the phase space point operators $A_\Omega^\gamma$ are vertices of the $\Lambda$ polytopes, and thus define phase space points of the $\Lambda$ polytope models. Here we consider a similar relation between the odd-prime-dimensional qudit extension of the CNC construction and the $\Lambda$ polytope models.

Let $\mathcal{S}$ denote the set of pure $n$-qudit stabilizer states. The $\Lambda$ polytope models are based on the set
\begin{equation}\label{eq:LambdaDef}
	\Lambda=\{X\in\Herm_1(d^n)\;|\;\Tr(\ket{\sigma}\bra{\sigma}X)\ge0\;\;\forall\ket{\sigma}\in\mathcal{S}\}.
\end{equation}
The set $\Lambda$ is a polytope for any dimension $d$ (see \cite[Lemma 1]{ZurelHeimendahl2024}). One can easily verify that $ A_\Omega^\gamma \in \Lambda $ for every $ A_\Omega^\gamma $ of the form Eq.~\eqref{eq:CNCPhasePointOperator} (see the proof of Lemma~\ref{Lemma:PauliUpdate}). More generally, we want to specify Hermitian operators in $ \Lambda $ whose expansion coefficients in the generalized Pauli basis are \emph{arbitrary} coefficients on the complex unit circle. That is, we want to consider Hermitian operators of the form 
\begin{equation}
	A_\Omega^\eta = \frac{1}{d^n}\sum_{u \in \Omega} \omega^{\eta(u)} T_u, \quad \Omega \subset E, \; \eta: \Omega \to [0,\pi] 
\end{equation}
that are contained in $ \Lambda $.
As it will turn out, the condition $ A_\Omega^\eta \in \Lambda$ implies that $ A_\Omega^\eta $ is precisely of the form \eqref{eq:CNCPhasePointOperator}, i.e. $ \Omega $ is closed under inference, $\eta(u) \pmod d \in \Z_d$, and
\begin{equation}
	\eta(a+b) \equiv \eta(a) + \eta(b) \pmod d \text{  if } [a,b] = 0.
\end{equation}

For a linear subspace $ L \subset E $ let 
\begin{equation}
	L^* := \{  \gamma: L \to \Z_d, \gamma(a+b) = \gamma(a)+\gamma(b)  \}	
\end{equation}
be its dual space, i.e. the space of linear functions on $ L $.

\begin{Theorem}\label{theorem:shape-A_omega-gamma}
	If $ A_\Omega^\eta \in \Lambda$, then 
	\begin{enumerate}
		\item [(i)] If $ I $ is an isotropic subspace and $ I \subset \Omega $, then $ \eta_{|I} \in I^* $.
		\item [(ii)]The set $ \Omega $ is closed under inference, i.e., if $ a,b \in \Omega $ and $ [a,b] = 0 $, then $ a+b \in \Omega $. 
	\end{enumerate}
\end{Theorem}

For $ M \subset E $ let $ \text{pr}_{M} : \Herm(d^n) \to \text{span} \{ T_b \, : \, b \in M \}$ be the projection that acts via
\begin{equation}
	\sum_{u \in E_n} c_b T_b \overset{\text{pr}_M}{\longmapsto} \sum_{b \in M} c_b T_b.
\end{equation}
To prove the lemma, we will use the fact that if $ X \in \Lambda $, then $ \text{pr}_I(X) \in \text{pr}_I(\Lambda) $.
Lemma~9 in \cite{ZurelHeimendahl2024} allows us to add redundant inequalities to $ \Lambda $, so that we can write 
\begin{equation}
	\Lambda=\{X\in\Herm_1(d^n)\;|\;\Tr(\Pi_I^\gamma X)\ge0\;\forall I, \, \forall\gamma \in I^*\}.
\end{equation}
where we range over all isotropic subspaces $I$ and all $\gamma\in I^*$. We will be mainly interested in $ \text{pr}_I(\Lambda) $ for $ I $ being an isotropic subspace. 

\begin{Lemma}\label{lemma:iso-subspace-projection}
	If $ I $ is an isotropic subspace, then $ \pr_I(\Lambda) $ is a self dual simplex with vertices $ \Pi_I^\gamma, \gamma \in I^* $, in particular $\textup{pr}_I(\Lambda) = \textup{conv} \{ \Pi_I^\gamma \, | \, \gamma \in I^* \}$.
\end{Lemma}

\emph{Proof of Lemma~\ref{lemma:iso-subspace-projection}.} Due to $ \text{pr}_I(\Pi_I^\gamma) = \Pi_I^\gamma $ it follows that  
\begin{equation}\label{eq:Lambda-isotropic-projection}
	\text{pr}_I(\Lambda) \subseteq \{ X = \sum_{b \in I} c_b T_b \, | \, \Tr(X) = 1, \, \Tr(X \Pi_I^\gamma) \ge 0  \, \text{ for all } \gamma \in I^* \}. 
\end{equation}
Analogously, to Lemma~15 in~\cite{ObstGross2024}, character orthogonality implies that the right hand side of~\eqref{eq:Lambda-isotropic-projection} is a self-dual simplex, so
\begin{align}
	\{ X =& \sum_{b \in I} c_b T_b \, | \, \Tr(X) = 1, \, \Tr(X \Pi_I^\gamma) \ge 0  \, \text{ for all } \gamma \in I^* \} \\
	=& \text{conv} \{ \Pi_I^\gamma \, | \, \gamma \in I^* \}
\end{align} 
and the vertices of $ \text{conv} \{ \Pi_I^\gamma \, | \, \gamma\in I^* \} $ are precisely the operators $ \Pi_I^\gamma $. On the other hand $ \Pi_I^\gamma \in \Lambda $ and therefore $ \Pi_I^\gamma \in \text{pr}_I(\Lambda)  $ for all $\gamma \in I^*$ and therefore also $ \textup{conv} \{ \Pi_I^\gamma \, | \, \gamma \in I^* \} \subseteq \textup{pr}_I(\Lambda) $. $\Box$

We will use the characterization Lemma~\ref{lemma:iso-subspace-projection} to prove Lemma~\ref{theorem:shape-A_omega-gamma}.
The overall strategy will be to argue that if $ \Omega $ and $ \eta $ violate one of the conditions of the theorem, then there is an isotropic subspace $ I \subset E$ such that $ \pr_I(A_\Omega^\gamma) \notin \pr_I(\Lambda)  $ implying $ A_\Omega^\gamma \notin \Lambda $. To show $ \pr_I(A_\Omega^\gamma) \notin \pr_I(\Lambda)  $, we will construct a hyperplane that separates the point $  \pr_I(A_\Omega^\gamma)$ from the polytope $\pr_I(\Lambda)$, that is, we construct some explicit Hermitian operator $Y\in \text{span}\{T_b\,|\, b\in I\} $ such that 
\begin{equation}\label{eq:hyperplane-separation-iso-projection}
	\Tr(\pr_I(A_\Omega^\gamma)Y) > \max_{X \in \pr_I(\Lambda)} \Tr(XY)  = \max_{\gamma\in I^*} \Tr(\Pi_I^\gamma Y), 
\end{equation}
where the last equation is a consequence of Lemma~\ref{lemma:iso-subspace-projection}.

\emph{Proof of Theorem~\ref{theorem:shape-A_omega-gamma}.} Let $ A_\Omega^\eta = \frac{1}{d^n}\sum_{u \in \Omega} \omega^{\eta(u)} T_u $ be Hermitian with $ \eta: \Omega \to [0,\pi] $ and $ \Tr(A_\Omega^\eta) =1 $ (the last condition implies $ 0 \in \Omega $ and $ \eta(0) = 0 $). As $ A_\Omega^\eta $ is Hermitian, we have 
\begin{equation}
	\sum_{u \in \Omega} \omega^{\eta(u)} T_u = A_\Omega^\eta = (A_\Omega^\eta)^\dagger = \sum_{u \in \Omega} \omega^{-\eta(u)} T_{-u},
\end{equation}
implying $ -\eta(u) = \eta(-u)$ for all $ u \in \Omega $. 

(i) Let $ I $ be an isotropic subspace contained in $ \Omega $. We will show that if $ A_\Omega^\gamma \in \Lambda $, then $ \eta_{|I} \equiv \gamma $ for some linear function $ \gamma: I \to \Z_d $: 
Therefore, assume that $ \eta_{|I} \not \equiv \gamma$ for all linear $ \gamma \in I^*$. In the sense of Equation~\eqref{eq:hyperplane-separation-iso-projection}, we can separate $ \textup{pr}_I(\Lambda) $ and $   \text{pr}_I(A_\Omega^\eta)$ by the hyperplane with normal vector $ A_I^{\eta_{|I}} \in \text{span} \{ T_b \, | \, b \in I  \}$:
\begin{align}
	\Tr(A_I^{\eta_{|I}} \text{pr}_I(A_\Omega^\eta)) =& \Tr(A_I^{\eta_{|I}} A_I^{\eta_{|I}} )\\
	=&\frac{1}{d^{2n}}\sum_{u \in I} \omega^{\eta(u)-\eta(u)} \Tr(T_uT_{-u})\\
	=&\frac{|I|}{d^n} > \Tr(A_I^{\eta_{|I}} \Pi_I^{\gamma}), 
\end{align}
where the strict inequality holds for all $ \gamma:I \to \Z_d $ with $ \gamma \neq \eta_{|I} $, so in particular for all $ \gamma \in I^*$.

(ii) Now suppose that $ \eta_{|I} \in I^*$ for any isotropic subspace $ I $ contained in $ \Omega $.  For (ii) assume that there are $ a,b \in \Omega $ with $ [a,b] = 0 $ such that $ a+b \notin \Omega $. We may assume that $ b \neq -a $ since $ 0 \in \Omega$.
Set $ I = \langle a,b \rangle $ (note that $ I \nsubseteq \Omega $) and 
\begin{align}
	Y =& \omega^{\eta(a)} T_a + \omega^{\eta(b)} T_b + \omega^{-\eta(a)} T_{-a} + \omega^{-\eta(b)} T_{-b} + \omega^{\eta(a)+ \eta(b) +\lfloor d/2 \rfloor } T_{a+b} + \omega^{-(\eta(a) +\eta(b) + \lfloor d/2 \rfloor )} T_{-(a+b)}\\
	&\in \text{span} \{T_x \, | \, x \in I \}. 
\end{align}
Then
\begin{equation}
	\Tr(\pr_I(A_\Omega^\gamma) Y) = 4 
\end{equation}
but for any $ \Pi_I^\gamma $ with $ \gamma \in I^* $
\begin{align}
	\Tr( \Pi_I^\gamma Y) =& (\omega^{\eta(a)-\gamma(a)}+\omega^{-\eta(a)+\gamma(a)} ) + (\omega^{\eta(b)-\gamma(b)}+\omega^{-\eta(b)+\gamma(b)} )\\
	&+ (\omega^{\eta(a)+\eta(b)-\gamma(a)-\gamma(b)+\lfloor d/2 \rfloor }+ \omega^{-(\eta(a)+\eta(b)-\gamma(a)-\gamma(b)+\lfloor d/2 \rfloor )} ) \\
	=&2\cos(\underbrace{\frac{2\pi (\eta(a)-\gamma(a))  }{d}}_{=:x}) + 2\cos(\underbrace{\frac{2\pi (\eta(b)-\gamma(b))  }{d}}_{=:y})\\
	& + 2\cos(\frac{2\pi (\eta(a)-\gamma(a)+ \eta(b)-\gamma(b)+\lfloor d/2 \rfloor )  }{d})   \\
	&= 2\left (\cos(x)+ \cos(y)+\cos(x+y+2\pi \frac{\lfloor d/2 \rfloor}{d}) \right )   \\
	&< 4.
\end{align}
This proves the theorem. $\Box$

\section{Conclusion}\label{Section:Conclusion}

In this work, we have presented the generalization of the CNC construction of Ref.~\cite{RaussendorfZurel2020} to the setting of odd-prime-dimensional qudits. We provide a characterization of the CNC phase space in this setting, and we describe its relation to other models like the Wigner function~\cite{Gross2006,Gross20062,Gross2008,VeitchEmerson2012} and the $\Lambda$ polytope models~\cite{ZurelRaussendorf2020,ZurelHeimendahl2024}. The phase space of this model contains the phase space of the Wigner function, but it also includes new phase space points which cannot be described as convex mixtures of Wigner function phase space points. We show that all vertices of the $\Lambda$ polytopes with coefficients of absolute value equal to one when expanded in the Pauli basis are CNC-type phase space point operators.

We also introduce a classical simulation algorithm for quantum computation with magic states based on sampling from probability distributions over the CNC phase space. Since the CNC construction outperforms the Wigner function and stabilizer methods in terms of the volume of states that can be positively represented, this new method allows a broader class of magic state quantum circuits to be efficiently classically simulated.

To conclude, we make two additional remarks commenting on possible avenues of future work.
\begin{enumerate}
	\item The CNC construction applies to qudits of any Hilbert space dimension, but the proof of efficiency of the update rules under Clifford gates and Pauli measurements relies on the characterization of the CNC phase space given in Ref.~\cite[\S IV]{RaussendorfZurel2020} for qubits and in Section~\ref{Section:PhaseSpaceCharacterization} for odd-prime-dimensional qudits. A similar characterization may be possible for qudits with composite Hilbert space dimension, but is likely much more complicated, in part because the set of labels $E:=\mathbb{Z}_d^{2n}$ of the Pauli operators fails to be a vector space for composite Hilbert space dimensions $d$~\cite{Gross2006}.
	\item For each fixed number $n$ of qudits, the volume of states that are positively represented by the CNC construction is strictly larger than the corresponding volume for the Wigner function. That said, talking about scaling of simulation cost with the number of qudits is more difficult here. The reason is the following: the phase space point operators of the Wigner function are closed under tensor products, and in fact, every phase space point operator can be constructed as a tensor product of single-qudit phase space point operators. As a result, the Wigner function is multiplicative: $W_{\rho\otimes\sigma}(u\otimes v)=W_\rho(u)W_\sigma(v)$ (see e.g. Ref.~\cite{VeitchEmerson2012}). When the representation goes negative for some state, the amount of negative as measured by the $1$-norm of the Wigner function is multiplicative in the number of copies of that state. For the case of stabilizer quasimixtures~\cite{HowardCampbell2017}, a tensor product of stabilizer states is again a stabilizer state, but there are other entangled stabilizer states that cannot be constructed in this way. As a result, the corresponding measure of negativity, the robustness of magic, is submultiplicative. On the other hand, for the CNC construction, a tensor product of CNC phase space point operators is generally not CNC. Thus, the equivalent of the robustness of magic for the CNC construction, another measure of negativity, is super-multiplicative. For example, two copies of the magic state $\ket{H}$ (see Figure~\ref{Figure:MagicTGate}) are nonnegative, but three copies are not~\cite{RaussendorfZurel2020}. This means we cannot obtain upper bounds on the negativity of product states by computing the negativity in the representation for few qudits. There are some tricks for partially addressing this, for example see Ref.~\cite[Appendix~D]{RaussendorfZurel2020} and~\cite{OkayRaussendorf2021}, but it remains more difficult to talk about scaling of quantum computational resources with the number of qudits for the CNC construction than for previous methods. We regard this as an open problem.
\end{enumerate}

The second issue above is also an present in other models derived from the $\Lambda$ polytopes, e.g.~\cite{ZurelRaussendorf2023}. While the $\Lambda$ polytopes are beginning to be explored for multiqubit systems~\cite{Heimendahl2019,RaussendorfZurel2020,OkayRaussendorf2021,OkayIpek2023,IpekOkay2023}, the extension of the $\Lambda$ polytope models to higher dimensional qudits is relatively new~\cite{ZurelHeimendahl2024}. Here we provide some initial structural insights, by characterizing the vertices with coefficients with absolute value equal to one in the Pauli basis, but we believe there is much more to be learned here.

In other settings, the odd-prime-dimensional qudit setting is simpler than the even-dimensional case, and was explored first. This is true for example for the link between Wigner functions and quantum computation~\cite{Galvao2005,CormickGalvaoGottesman2006,Gross2008,VeitchEmerson2012}. Following this pattern, we may be able to continue this work and derive insights from the $\Lambda$ polytopes for the simpler case of odd-prime-dimensional qudits which can later be imported to the case of qubits.

\section*{Acknowledgements}
M.Z. is funded by NSERC, in part through the Canada First Research Excellence Fund, Quantum Materials and Future Technologies Program. During the process of writing this work A.H. was partially supported by the Deutsche Forschungsgemein-schaft (DFG, German Research Foundation) under the Priority Program Compressed Sensing in Information Processing (CoSIP, project number SPP1798) and under Germany’s Excellence Strategy --- Cluster of Excellence Matter and Light for Quantum Computing (ML4Q) EXC 2004/1 - 390534769. This work was also supported by the Horizon Europe project FoQaCiA, GA no. 101070558.

\printbibliography

\appendix

\section{Proofs of lemmas~\ref{lemma:Adding-Element-to-subspace} and~\ref{lemma:Adding-element-to-cnc-set}}\label{Section:Proofs}

To prove the lemmas, we will require some additional definitions and statements. For a set $\Omega\subset E$, we define the (undirected) \emph{orthogonality graph} of $\Omega$ to be the graph $G(\Omega)=(\Omega,E)$ with vertex set $\Omega$ and edge set $E=\{\{a,b\}\;|\;[a,b]=0\}$, i.e., vertices in $G(\Omega)$ are adjacent if and only if the corresponding elements of $\Omega$ are orthogonal.

One crucial observation is the following. Suppose you are given a set $\Omega=\{a,b,c,d\}\subset E$ such that the orthogonality graph $G(\Omega)$ is a $4$-cycle (see Figure~\ref{Figure:4CycleOrthoGraph}). If the qudit Hilbert space dimension is $d=2$, i.e., the underlying field is $\mathbb{Z}_2$, then the orthogonal closure $\overline{\Omega}$ is the Mermin square~\cite{Heimendahl2019,KirbyLove2019,RaussendorfZurel2020}. In contrast, if the underlying field is $\mathbb{Z}_d$ and $d$ is an odd prime, then $\overline{\Omega}$ is the $4$-dimensional subspace spanned by $\{a,b,c,d\}$. This is the result of the following lemma.

\begin{Lemma}\label{lemma:comlpetion-of-4-cycle}
	If $\Omega=\{a,b,c,d\}$ and $G(\Omega)$ is a $ 4 $-cycle, then $\overline{\Omega}=\Span{a,b,c,d}$.
\end{Lemma}

The proof of this Lemma is based on the proof of Lemma~1 in Ref.~\cite{DelfosseRaussendorf2017}.

\emph{Proof of Lemma~\ref{lemma:comlpetion-of-4-cycle}.} The orthogonality relations are depicted in Figure~\ref{Figure:4CycleOrthoGraph}. Since $\Span{a},\Span{b},\Span{c},\Span{d}\subset\overline{\Omega}$, without loss of generality we may assume that $[a,c]=[b,d]=1$. Any point $v\in\Span{a,b,c,d}$ can be written as 
\begin{equation}
	v=(\alpha a+\delta d)+(\beta b+\gamma c)
\end{equation}
with $\alpha,\beta,\gamma,\delta\in\mathbb{Z}_d$. Then since $[a,b]=[a,c]=[b,d]=[c,d]=0$, it holds that $[\alpha a+\delta d,\beta b+\gamma c]=0$, and so it suffices to show that the planes spanned by non-orthogonal elements are contained in $\overline{\Omega}$, i.e., $\Span{a,d},\Span{b,c}\subset\overline{\Omega}$. But this follows immediately from the arguments in the proof of Lemma~1 in Ref.~\cite{DelfosseRaussendorf2017}. $\Box$

\begin{figure}
	\centering
	\begin{subfigure}{0.19\textwidth}
		\centering
		\begin{tikzpicture}
			\node at (0,2) (1) {$a$};
			\node at (2,2) (2) {$b$};
			\node at (0,0) (3) {$c$};
			\node at (2,0) (4) {$d$};
			\path (1) edge node{} (2);
			\path (1) edge node{} (3);
			\path (2) edge node{} (4);
			\path (3) edge node{} (4);
		\end{tikzpicture}
		\caption{\label{Figure:4CycleOrthoGraph}}
	\end{subfigure}
	\begin{subfigure}{0.19\textwidth}
		\centering
		\begin{tikzpicture}
			\node at (0,2) (1) {$a$};
			\node at (2,2) (2) {$b$};
			\node at (0,0) (3) {$c$};
			\node at (2,0) (4) {$d$};
			\path (1) edge node{} (2);
			\path (3) edge node{} (4);
		\end{tikzpicture}
		\caption{\label{Figure:2x2PathOrthoGraph}}
	\end{subfigure}
	\begin{subfigure}{0.19\textwidth}
		\centering
		\begin{tikzpicture}
			\node at (0,2) (1) {$a$};
			\node at (2,2) (2) {$b$};
			\node at (0,0) (3) {$x$};
			\node at (2,0) (4) {$d$};
			\path (1) edge node{} (2);
			\path (3) edge node{} (4);
			\path (1) edge node{} (3);
			\path (2) edge node{} (3);
		\end{tikzpicture}
		\caption{\label{Figure:LemmaOrthoGraphCase1a}}
	\end{subfigure}
	\begin{subfigure}{0.19\textwidth}
		\centering
		\begin{tikzpicture}
			\node at (0,2) (1) {$y$};
			\node at (2,2) (2) {$b$};
			\node at (0,0) (3) {$x$};
			\node at (2,0) (4) {$d$};
			\path (1) edge node{} (2);
			\path (3) edge node{} (4);
			\path (1) edge node{} (3);
			\path (2) edge node{} (3);
			\path (1) edge node{} (4);
		\end{tikzpicture}
		\caption{\label{Figure:LemmaOrthoGraphCase1b}}
	\end{subfigure}
	\begin{subfigure}{0.19\textwidth}
		\centering
		\begin{tikzpicture}
			\node at (0,2) (1) {$a$};
			\node at (2,2) (2) {$b$};
			\node at (0,0) (3) {$x$};
			\node at (2,0) (4) {$d$};
			\path (1) edge node{} (2);
			\path (3) edge node{} (4);
			\path (1) edge node{} (3);
		\end{tikzpicture}
		\caption{\label{Figure:LemmaOrthoGraphCase2}}
	\end{subfigure}
	\caption{Orthogonality graphs used in the proofs of Lemma~\ref{lemma:comlpetion-of-4-cycle}, and Lemma~\ref{prop:Extension-to-4-cycle-or-cnc-set}.}
\end{figure}
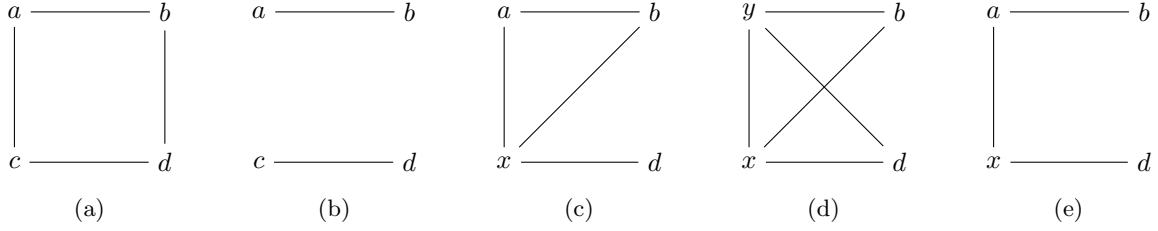

We also need the following additional lemma.

\begin{Lemma}\label{prop:Extension-to-4-cycle-or-cnc-set}
	Assume that $\Omega=\{a,b,c,d\}$ such that $\dim(\Span{a,b})=\dim(\Span{c,d})=2$ and $G(\Omega)$ is given by Figure~\ref{Figure:2x2PathOrthoGraph}. Then $\overline{\Omega}$ has the form Eq.~\eqref{eq:shape-of-cnc-sets}, or $\overline{\Omega}$ contains a subset $M$ such that $G(M)$ is a $4$-cycle.
\end{Lemma}

\emph{Proof of Lemma~\ref{prop:Extension-to-4-cycle-or-cnc-set}.} Since $ \dim(\langle c,d \rangle ) = 2 $, there is a nontrivial $x\in\Span{c,d}\subset\overline{\Omega}$, such that $[a,x]=0$. Now we distinguish two cases. For the first case, assume $[b,x]=0$, and so the orthogonality graph of $\{a,b,x,d\}$ is Figure~\ref{Figure:LemmaOrthoGraphCase1a}. Then there is a nontrivial $y\in\Span{a,x}\subset\overline{\Omega}$ such that $[d,y]=0$, and defining $I:=\Span{x,y}$ we have $b,d\in I^\perp$ (see Figure~\ref{Figure:LemmaOrthoGraphCase1b} for the orthogonality relations). Then the closure of $ \Omega $ is given by 
\begin{equation}
	\overline{\Omega}=\Span{b,I}\cup\Span{d,I}. 
\end{equation}

In the second case, assume $[b,x]\neq0$ (see Figure~\ref{Figure:LemmaOrthoGraphCase2} for the relevant orthogonality relations). Since $\dim(\Span{x,d})=2$ there is a nontrivial $y\in\Span{x,d}=\Span{c,d}$ such that $[b,y]=0$. Furthermore, as $a^\perp\cap\Span{c,d}=\Span{x}$, it holds that $[a,y]\neq0$. Then for $\Omega'=\{a,b,x,y\}$ we have $\overline{\Omega}=\overline{\Omega'}$ and the graph $G(\Omega')$ is a $4$-cycle. $\Box$

\medskip

\emph{Proof of Lemma \ref{lemma:Adding-Element-to-subspace}.} Let $\Omega\subset E$ be a subspace. If $v\in\Omega$, there is nothing to show, so assume $v\notin\Omega$. Clearly, 
\begin{equation}\label{eq:inclusion-orth-complement}
	\Span{v,\Omega\cap v^\perp}\subseteq\overline{\Omega \cup \{v \}}
\end{equation}
and if $\Omega\subset v^\perp$ we have equality in~\eqref{eq:inclusion-orth-complement} and $\overline{\Omega \cup\{v\}}$ is a subspace. Now assume that there is an $x\in\Omega$ such that $[v,x]\neq0$. Since $\dim(\Omega \cap v^\perp)\ge\dim(\Omega)-1$, it follows that $\Omega=\Span{x,\Omega\cap v^\perp}$.
Now we consider two cases:

(I)~If $\Omega\cap v^\perp$ is an isotropic subspace, then $H:=\langle v,\Omega\cap v^\perp\rangle$ is isotropic as well. Then $\dim(H\cap x^\perp)=\dim(H)-1$ and $ H \cap x^\perp \subset v^\perp $ and we can write the closure of $ \Omega $ as 
\begin{equation}
	\overline{\Omega} = \Span{x,H\cap x^\perp}\cup\Span{v,H \cap x^\perp}
\end{equation}
so $\overline{\Omega}$ has the form of Eq.~\eqref{eq:shape-of-cnc-sets}.

(II)~If $\Omega\cap v^\perp$ is not an isotropic subspace, then there are $a,b\in\Omega\cap v^\perp$ such that $[a,b]\neq0$. If $ a,b \in x^\perp $, then $ G(\{a,b,v,x\}) $ is a $ 4 $-cycle (see Figure~\ref{Figure:Lemma2OrthoGraphA}), and we apply Lemma~\ref{lemma:comlpetion-of-4-cycle}. Otherwise, since $\dim(\Span{a,b})=2$ we may assume without loss of generality that $[a,x]=0$ and $[b,x]\neq0$ (Figure~\ref{Figure:Lemma2OrthoGraphB}). Then we are precisely in case (2) of the proof of Lemma~\ref{prop:Extension-to-4-cycle-or-cnc-set} implying that $ \overline{\{a,b,v,x\}}=\Span{a,b,v,x}$. Consequently, $\Span{v,x}\subset\overline{\Omega\cup\{v\}}$. As $x$ was chosen arbitrarily in $\Omega\setminus v^\perp$, it follows $\overline{\Omega\cup\{v\}}=\Span{\Omega,v}$. $\Box$

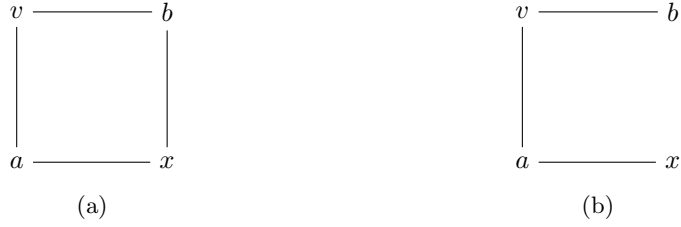
\begin{figure}
	\centering
	\begin{subfigure}{0.4\textwidth}
		\centering
		\begin{tikzpicture}
			\node at (0,2) (1) {$v$};
			\node at (2,2) (2) {$b$};
			\node at (0,0) (3) {$a$};
			\node at (2,0) (4) {$ x $};
			\path (1) edge node{}  (2) edge node {} (3);
			\path (4) edge node{}  (3) edge node{} (2);
		\end{tikzpicture}
		\caption{\label{Figure:Lemma2OrthoGraphA}}
	\end{subfigure}
	\begin{subfigure}{0.4\textwidth}
		\centering
		\begin{tikzpicture}
			\node at (0,2) (1) {$v$};
			\node at (2,2) (2) {$b$};
			\node at (0,0) (3) {$a$};
			\node at (2,0) (4) {$x$};
			
			\path (1) edge node{}  (2) edge node {} (3);
			\path (3) edge node{}  (4);
		\end{tikzpicture}
		\caption{\label{Figure:Lemma2OrthoGraphB}}
	\end{subfigure}
	\caption{Orthogonality graphs used in the proof of Lemma~\ref{lemma:Adding-Element-to-subspace}.}
\end{figure}

Finally, to prove Lemma~\ref{lemma:Adding-element-to-cnc-set}, we need one more observation, which follows from Lemma~\ref{lemma:Adding-Element-to-subspace}.

\begin{Corollary}\label{corollary:add-element-to-4-cycle-yields-subspace}
	If $\Omega\subset E$ is a subspace and contains $a,b,c,d$ such that $G(\{a,b,c,d\})$ is a $4$-cycle (Figure~\ref{Figure:4CycleOrthoGraph}), then $\overline{\Omega\cup\{v\}}=\Span{\Omega,v}$ for all $v\in E$.
\end{Corollary}

\emph{Proof of Corollary~\ref{corollary:add-element-to-4-cycle-yields-subspace}.} It suffices to prove that $  \Omega \cap v^\perp $ is not isotropic because then case (II) in the proof of Lemma~\ref{lemma:Adding-Element-to-subspace} can be applied. Therefore, let $I\subset\Omega$ be an isotropic subspace. As the largest isotropic subspace contained in $\Span{a,b,c,d}$ has dimension $2$, it follows that $\dim(\Span{a,b,c,d}\cap I)\le2$. Since $\Span{a,b,c,d}$ is a $4$-dimensional subspace contained in $\Omega$ and $I\subset\Omega$, it follows $\dim(I)\le\dim(\Omega)-2$. However, $\Omega\cap v^\perp$ is a $(\dim(\Omega)-1)$-dimensional subspace, hence, $\Omega\cap v^\perp$ is not isotropic. $\Box$

\emph{Proof of Lemma \ref{lemma:Adding-element-to-cnc-set}.} Let 
\begin{equation}
	\Omega=\bigcup\limits_{k=1}^{\xi}\Span{a_k,I}
\end{equation} 
where $I$ is an isotropic subspace, $a_k\in I^\perp$, and $[a_i,a_j]\neq0$ for $i\neq j$. We assume that $\xi>1$, otherwise we are in the situation of Lemma~\ref{lemma:Adding-Element-to-subspace}. As before, there is nothing to show if $v\in\Omega$, so suppose $v\notin\Omega$. We have two cases. 

(I)~First, assume $I\subset v^\perp$. Here we have two subcases.

(I.1)~If $[v,a_k]\neq0$ for all $k\in\{1,\dots,\xi\}$, then set $a_{\xi+1}=v$ and we have
\begin{equation}
	\overline{\Omega\cup\{v\}}=\bigcup\limits_{k=1}^{\xi+1}\Span{a_k,I}.
\end{equation}

(I.2)~Otherwise, without loss of generality we can assume that $[a_1,v]=0$. In this case there is a nontrivial $\tilde{v}\in\Span{a_1,v}\subset\overline{\Omega}$ such that $[a_2,\tilde{v}]=0$, and we can replace $v$ by $\tilde{v}$. Here again we have two subcases.

(I.2.a)~If $[a_2,\tilde{v}]=\cdots=[a_\xi,\tilde{v}]=0$, then $\tilde{I}:=\Span{\tilde{v},I}$ is an isotropic subspace and
\begin{equation}
	\overline{\Omega\cup\{v\}}=\bigcup\limits_{k=1}^{\xi}\Span{a_k,\tilde{I}}.
\end{equation}

(I.2.b)
If $ [\tilde{v},a_k]\neq0$ for some $k\in\{3,\dots,\xi\}$, without loss of generality $[\tilde{v},a_3]=0$. Then there is a nontrivial $v'\in\Span{a_1,v}=\Span{a_1,\tilde{v}}$ such that $ [a_3,v']=0$. Since $[\tilde{v},a_2]=0$ and $v'=\alpha a_1+\beta\tilde{v}\in\Span{a_1,\tilde{v}}$ for any $\alpha,\beta\in \Z_d$, it follows that
\begin{equation}
	[v',a_2]=\beta[a_1,a_2]\neq0.
\end{equation}
As $\tilde{v},v'$ are contained in the isotropic subspace $\Span{a_1,v}$ the orthogonality graph $G(\{\tilde{v},v',a_2,a_3\})$ is given by Fig.~\ref{fig:Orthogonality-graphs-CNC-Proof}. 

Applying case (2) of Lemma~\ref{prop:Extension-to-4-cycle-or-cnc-set} and Lemma \ref{lemma:comlpetion-of-4-cycle} shows that
\begin{equation}
	\Span{\tilde{v},v',a_2,a_3}= \overline{\{\tilde{v},v',a_2,a_3\}}\subset \overline{\Omega\cup\{v\}},
\end{equation}
and therefore, since $\tilde{v},v',a_2,a_3\in I^\perp$, the subspace $U'=\Span{I,\tilde{v},v',a_2,a_3}$ is contained in $\overline{\Omega}$. 

To conclude this subcase, observe that 
\begin{equation}
	\overline{\Omega}=\overline{U'\cup\{a_1\}\cup\{a_4\}\cup\cdots\cup\{a_\xi\}}.
\end{equation}
Now consider $\overline{U'\cup \{h_1\}}$ and use the fact that we can construct a $4$-cycle in $\Span{\tilde{v},v',h_2,h_3}\subset U'$. Thus, we are able to apply Corollary~\ref{corollary:add-element-to-4-cycle-yields-subspace} to get $\overline{U'\cup\{a_1\}}=\Span{U', a_1}$ and iteratively we obtain
\begin{equation}
	\overline{\Omega}=\overline{U'\cup\{a_1\}\cup\{a_4\}\cdots\cup\{a_\xi\}}=\Span{I,a_1,\dots,a_\xi,v}.
\end{equation}

(II)~For the second case, suppose that $I\nsubseteq v^\perp$. Let $u\in I$ with $[u,v]\neq0$. Since  $a_1,a_2\in I^\perp\subset u^\perp$ we can find a nontrivial $\tilde{a}_1\in\Span{u,a_1}$ and a $\tilde{a}_2\in\Span{u,a_2}$ such that $[\tilde{a}_1,v]=[\tilde{a}_2,v]=0$. Since $[\tilde{a}_1,\tilde{a}_2]=[a_1,a_2]\neq0$, the orthogonality graph $G(u,v,\tilde{a}_1,\tilde{a}_2)$ is given in Figure~\ref{fig:Orthogonality-graphs-CNC-Proof6} and therefore a $4$-cycle. Hence, by Lemma \ref{lemma:comlpetion-of-4-cycle}, it follows that $\overline{\Omega\cup\{v\}}$ contains the subspace $\Span{\tilde{a}_1,\tilde{a}_2,u,v}$. Now we can iteratively add the remaining elements of $I$ and the cosets $a_3+I,\dots a_\xi+I$ to $\Span{\tilde{a}_1,\tilde{a}_2,u,v}$ and apply Corollary~\ref{corollary:add-element-to-4-cycle-yields-subspace} to obtain $\overline{\Omega}=\Span{I,a_1,\dots,a_\xi,v}$. $\Box$

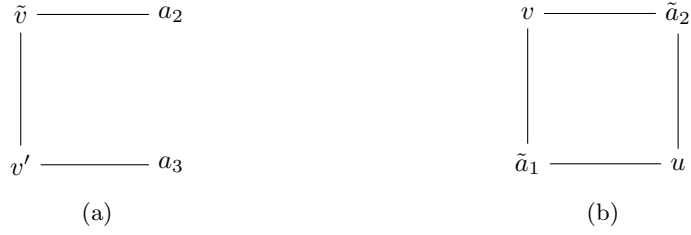
\begin{figure}
	\centering
	\begin{subfigure}{0.4\textwidth}
		\centering
		\begin{tikzpicture}
			\node at (0,2) (1) {$\tilde{v}$};
			\node at (2,2) (2) {$a_2$};
			\node at (0,0) (3) {$v'$};
			\node at (2,0) (4) {$ a_3$};
			\path (1) edge node{}  (2) edge node {} (3);
			\path (3) edge node{}  (4);
		\end{tikzpicture}
		\caption{\label{fig:Orthogonality-graphs-CNC-Proof}}
	\end{subfigure}
	\begin{subfigure}{0.4\textwidth}
		\centering
		\begin{tikzpicture}
			\node at (0,2) (1) {$v$};
			\node at (2,2) (2) {$\tilde{a}_2$};
			\node at (0,0) (3) {$\tilde{a}_1$};
			\node at (2,0) (4) {$ u$};
			\path (1) edge node{}  (2) edge node {} (3);
			\path (3) edge node{}  (4);
			\path (2) edge node{}  (4); 
		\end{tikzpicture}
		\caption{\label{fig:Orthogonality-graphs-CNC-Proof6}}
	\end{subfigure}
	\caption{Orthogonality graphs used in the proof of Lemma~\ref{lemma:Adding-element-to-cnc-set}.}
\end{figure}

\end{document}